\documentclass[fleqn,usenatbib]{mnras}

\usepackage{newtxtext,newtxmath}

\usepackage[T1]{fontenc}

\DeclareRobustCommand{\VAN}[3]{#2}
\let\VANthebibliography\thebibliography
\def\thebibliography{\DeclareRobustCommand{\VAN}[3]{##3}\VANthebibliography}

\usepackage{graphicx}	
\usepackage{amsmath}	

\usepackage{xspace}
\usepackage{xcolor}
\usepackage{float}
\usepackage[normalem]{ulem}
\usepackage{orcidlink}

\newcommand{\degree}{{$^\circ$}\xspace}

\newcommand{\snr}{{SNR\textsubscript{15\AA{}}}\xspace}

\defcitealias{TiDES_2025}{CF}

\title[YSN Pipeline]{A Young Supernova Selection Pipeline For The LSST Era}

\author[H. Addison et al.]{Harry Addison $^{1}$\thanks{E-mail: ha00871@surrey.ac.uk} \orcidlink{0000-0001-8271-1364},
Chris Frohmaier$^{2}$ \orcidlink{0000-0001-9553-4723},
Kate Maguire$^{3}$ \orcidlink{0000-0002-9770-3508},
Robert C. Nichol$^{1}$,
Isobel Hook$^{4}$ \orcidlink{0000-0002-2960-978X},
\newauthor
Stephen J. Smartt$^{5, 6}$ \orcidlink{0000-0002-8229-1731
}
\\
$^{1}$Department of Physics, University of Surrey, Guildford GU2 7XH, United Kingdom \\
$^{2}$Institute of Cosmology and Gravitation, University of Portsmouth, Portsmouth PO1 3FX, UK \\
$^{3}$School of Physics, Trinity College Dublin, The University of Dublin, Dublin 2, Ireland \\
$^{4}$Department of Physics, Lancaster University, Lancaster, Lancashire LA1 4YB, UK \\
$^{5}$Department of Physics, University of Oxford, Keble Road, Oxford, OX1 3RH, UK \\
$^{6}$Astrophysics Research Centre, School of Mathematics and Physics, Queen’s University Belfast, BT7 1NN, UK
}

\date{Accepted XXX. Received YYY; in original form ZZZ}

\pubyear{2025}

\begin{document}
\label{firstpage}
\pagerange{\pageref{firstpage}--\pageref{lastpage}}
\maketitle

\begin{abstract}
Early-time spectroscopy of supernovae (SNe), acquired within days of explosion, yields crucial insights into their outermost ejecta layers, facilitating the study of their environments, progenitor systems, and explosion mechanisms. Recent efforts in early discovery and follow-up of SNe have shown the potential insights that can be gained from early-time spectra.    
Surveys such as the Time-Domain Extragalactic Survey (TiDES), conducted with the 4-meter Multi-Object Spectroscopic Telescope (4MOST), will provide spectroscopic follow-up of transients discovered by the Legacy Survey of Space and Time (LSST). Current simulations indicate that early-time spectroscopic studies conducted with TiDES data will be limited by the current SN selection criteria.
To enhance early-time SN spectroscopic studies from TiDES-like surveys, we propose a set of selection criteria focusing on young SNe (YSNe), which we define as SNe prior to $-10$ days before peak brightness. Utilising the Zwicky Transient Facility transient alerts, we developed criteria to select YSNe while minimising the sample's contamination rate to 23 percent.
The developed criteria were applied to LSST simulations, yielding a sample of 694 Deep Drilling Field survey SNe and 56260 Wide Fast Deep survey SNe for follow-up.
We demonstrate that our criteria enables the selection of SNe at early-times, enhancing future early-time spectroscopic SN studies from TiDES-like surveys.
Finally, we investigated 4MOST-like observing strategies to increase the sample of spectroscopically observed YSNe. We propose that a 4MOST-like observing strategy that follows LSST with a delay of 3 days is optimal for a TiDES-like SN survey in terms of the number of classifiable spectra obtained, while a 1 day delay is most optimal for enhancing the early-time science in conjunction with our YSN selection criteria.

\end{abstract}



\section{Introduction}

Supernovae (SNe) are a diverse set of transients with many different progenitor scenarios and explosion mechanisms. For the case of thermonuclear (Type Ia) supernovae, there are two traditional progenitor scenarios (single-degenerate scenario \citep{Whelan_1973} and double-degenerate scenario \citep{Iben_1974}), while there are many different explosion mechanisms being studied, which include
pure deflagration models \citep{Nomoto_1984, Jordan_2012, kromer_2013, Fink_2014, Lach_2022}, deflagration-to-detonation (delayed-detonation) models \citep{Arnett_1969, Khokhlov_1991, Gamezo_2005, Ropke_2007, Rabinak_2012, Seitenzahl_2013}, double-detonation models \citep{Fink_2007, Fink_2010, Kromer_2010, Shen_2014, Polin_2019, Magee_2021, Boos_2024}, core-degenerate explosions \citep{Soker_2011, Wang_2016}, triple collision models \citep{Kushnir_2013, Hallakoun_2019}, and rotating super-Chandrasekhar mass explosions \citep{Stefano_2011}.

To investigate the explosion mechanisms of SNe Ia, previous studies have investigated the early light curves of SN Ia events \citep{Riess_1999, Hayden_2010, Bianco_2011, Ganeshalingam_2011, Firth_2014, Miller_2020, Burke_2022a, Burke_2022b, Deckers_2022, Fausnaugh_2023, Ni_2025}. In addition to early-time photometry, early-time spectroscopy can further provide us with a wealth of information about the explosion dynamics. Spectra taken within a few days of explosion allow us to trace the outermost layers of the ejecta. From these spectra we can determine the chemical abundances of the outer layers, which can then be used to distinguish between explosion models \citep{Magee_2021,Ogawa_2023}.
Early-time spectra can also be used to study SNe Ia that display a ``bump'' in their pre-peak light curves, which is thought to be the result of the shock front interacting with the binary companion \citep{Kasen_2010}, or from the presence of short-lived radioactive isotopes \citep{Noebauer_2017}.

Furthermore, early-time spectra are not only useful for studying thermonuclear SN explosion mechanisms, but they can also be used to study core collapse (CC) SNe. Early-time spectra can be used to investigate the progenitor systems of CC SNe by providing us with the ability to constrain the progenitor's chemical abundance, wind speed, and mass loss \citep{pastorello_2007, Smith_2007, Smith_2023, Zimmerman_2024}. 
As with SNe Ia, some CC SNe also exhibit a ``bump'' in their pre-peak light curves, which is caused by an interaction of the shock front with circumstellar material \citep[CSM;][]{Piro_2016, Gagliano_2022, Kozyreva_2022}. Spectra taken within hours to days of explosion, known as flash spectroscopy, can reveal narrow emission lines from the shock breakout flash-ionisation and recombination of the CSM , offering a direct probe of the progenitor's immediate environment and final stages of mass loss \citep{gal-yam_2014, khazov_2016, Yaron_2017, Kochanek_2019, Bruch_2023, Jacobson-gal_2023, Zimmerman_2024}. 

Besides the study of SN progenitors and explosion mechanisms, early-time spectra can also be very useful for classifications. Some types of SNe, such as stripped envelope SN that retain a small hydrogen envelope, can only be reliably classified from their early phase spectra \citep{dong_2023}. Additionally, the ability to spectroscopically classify SNe early in their evolution allows us to conduct targeted observations of high interest SNe, such as the previously mentioned ``bump'' SNe or other peculiar types.   

Currently, our spectroscopic samples of SNe, and in particular early-time spectra, are restricted by the availability of spectroscopic resources that can quickly follow-up photometrically discovered transients. The {\sl Zwicky Transient Facility} \citep[ZTF;][]{ZTF_Bellm_2018, ZTF_Masci_2018} is one of the leading facilities of transient astronomy, with its Northern Sky Survey observing the northern sky (declination $> -31$\degree) every 2 days in both the $g$- and $r$-bands. This has allowed for the discovery of many transient events, however, a key part of ZTF's success is its {\sl Bright Transient Survey} \citep[BTS;][]{Fremling_2020}. BTS uses an automated low resolution integral field unit spectrograph, the Spectral Energy Distribution Machine \citep{ben-Ami_2012, Blagorodnova_2018, Rigault_2019}, to compliment the Northern Sky Survey by providing spectroscopic follow-up of the ZTF discovered transients. The main aim of BTS is to provide a complete sample of spectroscopically classified extra-galactic transients within the Northern Sky Survey that are brighter than 18.5 mag \citep{Fremling_2020}. Between June 2018 and April 2025, ZTF and BTS have found and spectroscopically classified over 10600 SNe\footnote{ZTF Bright Transient Survey, \url{https://sites.astro.caltech.edu/ztf/bts/bts.php} (accessed 13/06/2025)}. 

In the coming years the detection rates of transients will be greatly increased, with the Vera C. Rubin Observatory's {\sl Legacy Survey of Space and Time} \citep[LSST;][]{LSST_Ivezic_2019} discovering millions of SNe. Along with increased photometric observations of SNe, we will also see a large increase in our spectroscopic SN samples. The {\sl Time-Domain Extragalactic Survey} \citep[TiDES;][]{TiDES_2019, TiDES_2025} is one of the twenty five surveys to be conducted on the {\sl 4 metre Multi-Object Spectroscopic Telescope} \citep[4MOST;][]{4MOST_2019}, providing follow-up spectra of LSST transients. TiDES will operate over 5 years in parallel with the other 4MOST surveys, constructing the largest spectroscopic cosmological SN sample to date (\citet{TiDES_2025}, hereafter \citetalias{TiDES_2025}).

In order to obtain follow-up spectra of LSST transients, TiDES will select its targets by applying sets of selection criteria to the real time transient alerts that LSST will send out \citepalias{TiDES_2025}. The different sets of selection criteria will be used in conjunction with one another to enhance the scientific output of TiDES. The current proposed selection criteria for SN are described by \citetalias{TiDES_2025}, and are as follows:
\begin{itemize}
    \item Transient detected to $>5\sigma$ in three or more bands.
    \item Transient observed on two distinct nights.
    \item Transient is brighter than 22.5 mag in any $griz$ filter.
\end{itemize}
Using simulations of LSST and 4MOST, \citetalias{TiDES_2025} showed that their current SN selection criteria predominantly selects pre-peak SNe from the LSST alerts. However, they also showed that there is on average a seven-day delay between the selection of a target from LSST and its subsequent follow-up using 4MOST. This results in many of the obtained SN spectra being taken during post-peak SN phases, with very few spectra obtained for phases (relative to peak brightness hereafter) before $-15$ days. Therefore, studies such as that of SN explosion mechanisms, progenitor compositions, wind speeds, mass loss, and early-time CSM/binary companion interactions will be limited with the TiDES spectroscopic SN sample based on current plans.    

In this study, we propose a new set of selection criteria for TiDES-like surveys that are focused on selecting transients as early as possible for the purpose of producing a SN sample for early-time astrophysical studies. Our criteria are designed to be used in conjunction with other selection criteria that are implemented within TiDES-like surveys, with a specific focus on enhancing the sample of young (early-time) SN (YSN) spectra. Throughout this study we define a YSN as a SN at a phase prior to $-10$ days with respect to peak brightness. Whilst early selection is our primary objective, our selection criteria must also ensure that the YSN sample is not highly contaminated. This is due to the limited number of observing hours available for TiDES-like surveys that we do not want to waste on non-real sources. 
By using the ZTF transient alerts and LSST simulations, we aim to demonstrate that our proposed selection criteria will provide TiDES-like surveys with more early-time SN targets to follow-up than the current selection criteria, whilst minimising the contamination.

Additionally, we investigate observing strategies with the aim of optimising a 4MOST-like observing strategy for the use case of a TiDES-like SN survey and our YSN selection criteria. 
We explore different observing strategies to improve the quality of the obtained SN spectra, also considering the need for a quick follow-up of the targets in order to make full use of the early SN selection provided by our YSN selection criteria. We do not consider the impact that the investigated observing strategies have on the non-SN surveys when determining an optimal strategy.
In this work our objective is not to produce a full 4MOST-like observing strategy, but rather to provide guidelines for future works towards developing and simulating 4MOST-like observing strategies. 

This paper is organised as follows: in Section \ref{sec:ztf_development} we present the development of our YSN selection criteria using the ZTF transient alerts. 
In Section \ref{sec:lsst_application}, we adapt and apply our developed selection criteria to an LSST simulation, presenting the resulting YSN candidate sample. The YSN candidate sample is then evaluated by comparing it to the SN sample produced by the current TiDES selection criteria. 
In Section \ref{sec:optimal_strategy}, we investigate different 4MOST-like observing strategies, attempting to optimise the output of the TiDES-like surveys and our YSN selection criteria. 
Finally, in Section \ref{sec:conclusion} we summarise the conclusions of this study.


\section{YSN Selection Criteria Development}\label{sec:ztf_development}

Transient follow-up surveys such as TiDES will obtain targets from the live transient alerts that LSST will produce. As LSST is due to start operations in late 2025 \footnote{LSST project status, available at \url{https://www.lsst.org/about/project-status} (accessed 17/10/2024)}, we must look to other means of developing and testing our selection criteria.
One way in which this could be done is with the LSST simulations such as those used by \citetalias{TiDES_2025}, which simulate a SN population and their LSST photometric observations. However, one major limitation of these simulations is that they focus on the extragalactic Universe, excluding Galactic events (such as cataclysmic variables) that form a major source of contamination for a young and bright extragalactic transient search. 
Therefore, the LSST simulations used by \citetalias{TiDES_2025} do not fully represent the transient/variable sky that will be present in the live LSST transient alerts.
As we want to produce a high purity (low contamination) YSN sample, as to not waste fibre hours (observing time) on non-real sources, using the LSST simulations alone is not enough to investigate this requirement. Therefore, we made use of the ZTF transient alerts, which will be similar to the LSST alerts, to develop and investigate suitable selection criteria that produce a high purity YSN sample.

\subsection{Proposed YSN Selection Criteria}\label{subsec:proposed_selection_criteria}

ZTF has been observing the northern sky with a cadence of two days in the $g$- and $r$-filter bands since December 2020\footnote{ZTF public data release 5 notes, available at \url{https://irsa.ipac.caltech.edu/data/ZTF/docs/releases/dr05/ztf\_release\_notes\_dr05.pdf} (accessed on 28/04/2025)}. When a source is detected to vary above a specified detection threshold in the difference image, an alert is sent out to the wider community through brokers \citep{Patterson_2018}. These alerts contain information about the source including, but not limited to; its object ID, coordinates, magnitude and filter band, associated detections from the last 30 days, and cutouts of science and difference images. 
Additionally, brokers provide their own data products to enhance the ZTF alerts. For example, the Lasair broker \citep{lasair, Williams_2024} supplies a contextual classification of the object using its contextual classifier, Sherlock \citep{sherlock_Young}.     
In this study we adopted Lasair as our broker of choice, which is predominantly motivated by its use by TiDES for development purposes.

We considered the information provided in the ZTF alerts, along with additional data products provided by Lasair, and proposed a set of YSN selection criteria. These proposed criteria
are presented in Table \ref{tab:ztf_selection_criteria} along with a summary of the motivations
behind using them, which are discussed in more detail below.

\begin{table*}
    \caption{Developed YSN selection criteria for use on the ZTF transient alerts. Provided are the criteria along with a description of each criterion's purpose.
    \\ * This criterion has no effect when applied to ZTF transient alerts as ZTF only detects objects brighter than $21.5$ mag.\\ 
    ** The multiple values provided are the values that were tested, and only one value for a given criterion was applied at any given time.}
    \label{tab:ztf_selection_criteria}
    \centering
    \begin{tabular}{c|c}
        \hline
        \hline
            Criterion & Reasoning \\
        \hline
        \hline
            $g$- or $r$- band magnitude $<22.5$ mag * & Ensures that the object will meet the TiDES SN SSC \\
            Galactic latitude $<-10$\degree OR Galactic latitude $>10$\degree & Removal of Galactic transient sources \\
            Number of $g$- or $r$-band $>5 \sigma$ positive difference detections $\geq 2$ & Two or more $>5 \sigma$ $g$- or $r$-band detections required for brightening rate criterion \\
            \texttt{Sherlock} classification not: VS, AGN, CV, BS & Removal of contaminants \\
            Age $<$ 7, 14 days ** & Removes older objects that are unlikely pre-peak SNe \\
            Brightening rate $> 0.2$, $0.1$, $0.05$ mag/day ** & Removes dimming and slow brightening sources that are unlikely to be YSNe. \\ 
        \hline
        \hline 
    \end{tabular}
\end{table*}

Our first proposed criterion, objects must be brighter than 22.5 mag in either the $r$- or $g$-band, was adopted from the current TiDES selection criteria. \citetalias{TiDES_2025} used this to ensure that a selected object is bright enough to meet the TiDES SN spectral success criteria (SSC), which is a criterion that defines whether or not an observation was successful. The SSC for TiDES SN is that for a given spectrum the mean signal to noise ratio in the wavelength range $4500-8000$\AA\ is more than 5 per 15 \AA, which is based on the ability of SN classification tools to provide reliable classifications \citepalias{TiDES_2025}. Therefore, we also require our objects to be brighter than 22.5 mag. It is worth mentioning that ZTF has a best case $5\sigma$ limit of ${\sim}21.5$ mag \citep{ZTF_Bellm_2018}, meaning that this criterion has no effect on the produced ZTF samples. 

The second criterion, exclusion of objects between Galactic latitudes of $-10$\degree and $10$\degree, was used to remove the Galactic plane from our selection region. We are only interested in extragalactic transients. Therefore, this criteria reduces the number of Galactic transients that are selected by our selection criteria.

Our next criterion, an object must have two or more positive difference detections greater than $5\sigma$ in the same band, is to exclude objects with a single or no positive detections in both the $g$- and $r$-bands. As is made clear in Section \ref{subsec:ztf_application_methods}, by applying this criterion we reduced the computational resources and time required to apply our last two criteria.

Next we utilised \texttt{Sherlock} \citep{sherlock_Young}, a contextual classifier, to filter out objects that we considered as contamination such as: variable stars (VS), active galactic nuclei (AGN), cataclysmic variables (CV), and bright stars (BS). The classifier is not 100 percent accurate, but we are not looking for a complete sample of YSNe, and our last two selection criteria are focused on removing any contaminants that remain.    

Our penultimate criterion, referred to as the age constraint, is a constraint on the time since the first five sigma positive difference detection of an object. This was utilised to reduce the contamination of our selected YSN candidate sample by removing long-lived transients. The age constraint should also reduce the number of post peak SNe in the sample as it will remove SN that have been observed over longer periods of time, which are likely no longer early-time SNe.
Unlike the previous criteria, the age criterion does not have a clear definitive threshold value, so we tested two values: 7 and 14 days.

Our final criterion, a brightening rate constraint, was used to further remove contaminants from our YSN candidate sample. As the brightening rate of a SN in its early phases is generally much more rapid than at near-peak phases, we exploited this property to filter out ``slowly'' brightening and dimming sources that are unlikely to be linked to the early phases of a SN outburst. 
To employ this criterion, we used equation \ref{eqn:bright_rate} to calculate the brightening rate between the latest observation and the mean of the second latest night's observations of the same filter band. 
\begin{equation}\label{eqn:bright_rate}
    \mathrm{Brightening~rate} = \frac{- (m_{\rm{latest}} - m_{\rm{night 2}})}{\rm{JD}_{\rm{latest}} - \rm{JD}_{\rm{night 2}}}
\end{equation}
where the brightening rate is in units of magnitudes per day (mag/day), $m_{\rm{latest}}$ is the apparent magnitude of the latest observation, $m_{\rm{night 2}}$ is the mean magnitude of the previous available night's observations in the same filter band as the latest observation, $\rm{JD}_{\rm{latest}}$ is the Julian date (JD) of the latest observation, and $\rm{JD}_{\rm{night 2}}$ is the average JD of the previous available night's observations in the same filter band as the latest observation. We associate a positive brightening rate with a source that is brightening, hence the use of the negative magnitude difference between the two nights ($m_{\rm{latest}} - m_{\rm{night 2}}$) in equation \ref{eqn:bright_rate}.
We compared the latest observation to the observations from the previous available night, as observations (of the same band) taken within the same night can vary in magnitude (but within error), which in some cases can cause a false sense of brightening or dimming between the consecutive observations. For the same reason, we took the mean of the $g$- or $r$-band observations on the previous available night.

As with the age criterion, the brightening rate does not have a definitive threshold value, and so we applied and tested three values: $0.05$ mag/day, $0.1$ mag/day, and $0.2$ mag/day. For an object to be selected as a YSN, it had to have a brightening rate more positive than the mentioned values, which corresponds to a more rapid brightening than the threshold values.

\subsection{Querying Archived Alerts and Lasair for Transient Data}\label{subsec:ztf_application_methods}

In order to choose the most suitable values for the age and brightening rate criteria, as well as to evaluate our proposed YSN selection criteria, we produced YSN candidate samples by applying our criteria to the ZTF archived transient alerts\footnote{ZTF Alert Archive \url{https://ztf.uw.edu/alerts/public/} (accessed on 13/05/2025).}. We began by applying the magnitude, galactic latitude, and number of positive difference detection constraints to the alerts (the first three criteria in Table \ref{tab:ztf_selection_criteria}).

As the alerts do not contain enhancements by the brokers, to apply our Sherlock classification criteria, we queried Lasair for the Sherlock classifications of the objects that passed our initial filtering. This was achieved by using the table query method within Lasair's \texttt{Python} API, which allows you to perform an SQL query on the Lasair database. Objects with Sherlock classifications of VS, AGN, CV, or BS were removed from the sample.

Next we applied the age and brightening rate criteria by obtaining the light curves of the objects that remained in our sample.
We made use of the light curve query method within Lasair's \texttt{Python} API, which allows you to download the full light curve (detections and non-detections) of an object using its ZTF candidate ID. 
We then calculated the ``age'' of the object and applied the age constraint to it, producing an ``age'' restricted sample.  
For this sample of objects, the brightening rate was then calculated and the brightening rate criterion was applied to produce a YSN candidate sample.

\subsection{YSN Selection Criteria Testing}\label{subsec:criteria_testing}

Throughout the development of our YSN selection criteria, six different YSN candidate samples were produced to test the different proposed age and brightening rate threshold values. The details of how these YSN candidate samples were produced, along with discussions of these samples for the purpose of development and evaluation of our criteria are provided in the sections that follow.

\subsubsection{Producing ZTF YSN Candidate Samples}\label{subsubsec:produce_ztf_samples}

To select a suitable constraint values for the ambiguous age and brightening rate criteria, we first produced samples of objects by applying the methods outlined in Section \ref{subsec:ztf_application_methods} to the archived ZTF alerts. We selected archived alerts from a total of 60 nights split between a summer period (5$^{\rm{th}}$ June - 10$^{\rm{th}}$ July 2023) and a winter period (10$^{\rm{th}}$ December 2023 - 29$^{\rm{th}}$ January 2024). These periods exceed a total of 60 nights as we did not select nights where ZTF did not observe. We note here that we excluded observations from the nights of 13$^{\rm{th}}$, 14$^{\rm{th}}$, and 15$^{\rm{th}}$ December 2023, as ZTF was performing an extragalactic high cadence experiment\footnote{ZTF experiments, available at \url{https://www.ztf.caltech.edu/ztf-experiments.html}, (accessed on 21/05/2025)}, which is not representative of normal survey operations and could impact our results.   
As mentioned in Section \ref{subsec:proposed_selection_criteria}, we tested two threshold values (7 and 14 days) for the age criterion, and three threshold values (0.05, 0.1, and 0.2 mag/day) for the brightening rate criterion. Therefore, we produced six samples, each produced from a different combination of the age and brightening rate criteria threshold values.

\subsubsection{Analysing ZTF YSN Candidate Samples}\label{subsubsec:analyse_ztf_samples}

To select the more optimised constraint values for our needs, we investigated the contamination (non-YSN objects) of the samples. This was accomplished by cross matching the objects to the {\sl Transient Name Server} \citep[TNS;][]{TNS}, obtaining their classifications if possible. 
For the TNS classified SNe, we obtained the JD of peak brightness from the BTS sample explorer\footnote{BTS sample explorer, available at \url{https://sites.astro.caltech.edu/ztf/bts/explorer.php} (accessed on 13/05/2025)}. If a SN was not present in the BTS sample, then the JD of peak brightness was estimated by fitting a fourth degree polynomial to the $g$-band light curve if possible, otherwise the $r$-band light curve was used. The times of maximum from the polynomial fits have uncertainties in the range of one day to five days, depending on the light curve sampling near maximum.
The JD of peak brightness was used to calculate the phase at which a given SN was selected, and subsequently determine if it was a YSN (phase $<-10$ days), pre-peak SN (phase $<0$ days), or post-peak SN (phase $\geq0$ days) at the time of selection. It should be noted that YSNe are a subset of pre-peak SNe.

\begin{figure*}
    \begin{tabular}{ccc}
         \includegraphics[width=0.320\linewidth]{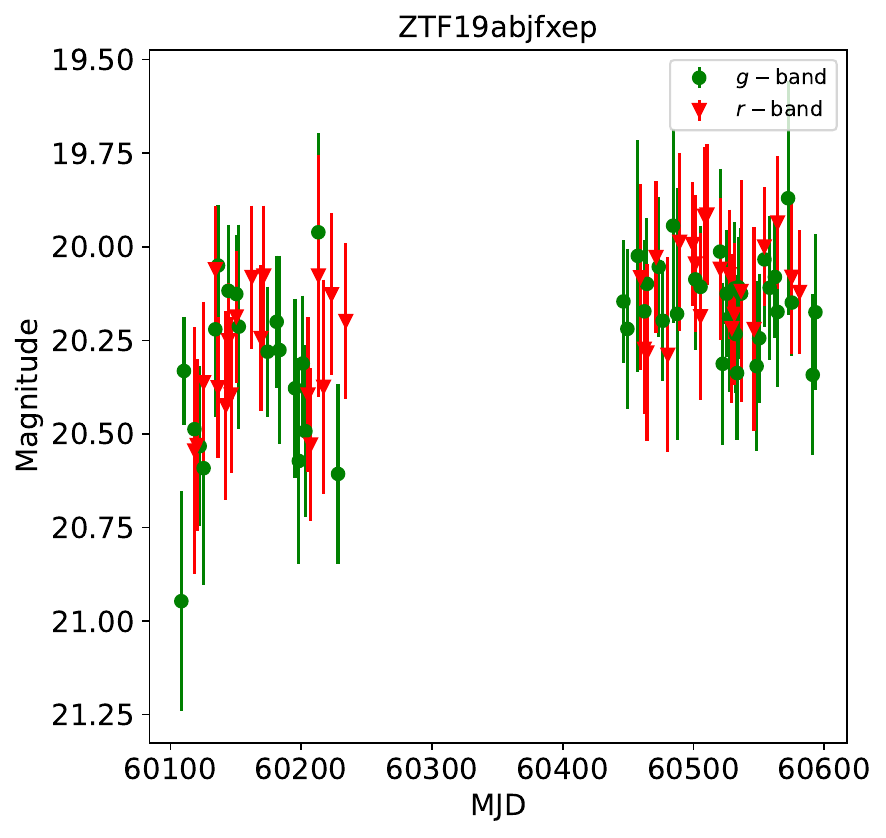} &
         \includegraphics[width=0.305\linewidth]{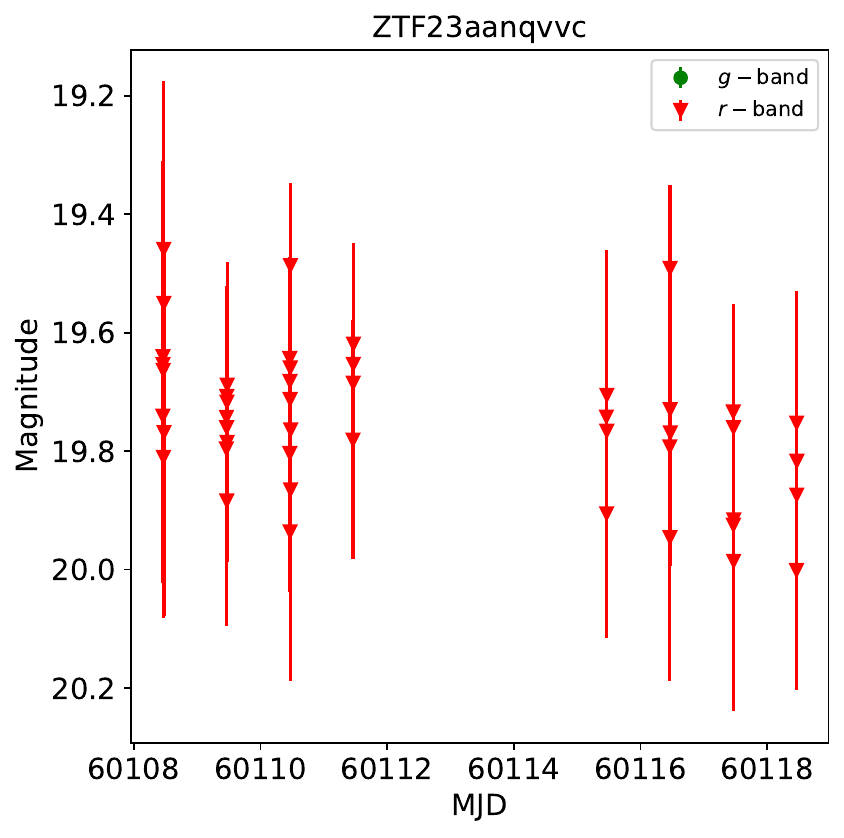} &
         \includegraphics[width=0.305\linewidth]{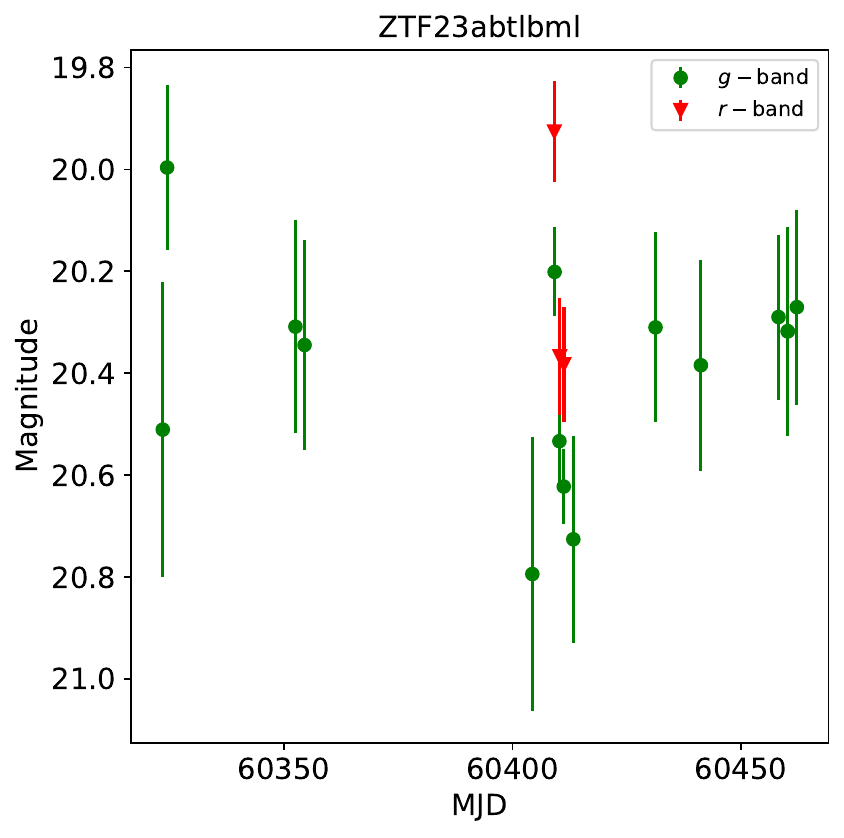}
    \end{tabular}
   
    \caption{Light curves of unclassified contaminants in our YSN candidate sample selected from the ZTF transient alerts.
    (Left) $g$- and $r$-band light curves of a contaminant that is likely an AGN or quasar. (Middle and right) $g$- and $r$-band light curves of contaminants that have no clear nature.}
    \label{fig:ztf_unclass_contam_lcs}
\end{figure*}

Additionally, we further investigated the nature of the unclassified objects, as some of these objects could be unclassified SNe, and possibly YSNe. We visually inspected all of the unclassified objects' light curves resulting from our different selection criteria (1025 total objects), considering their shape, rise, time-scales, and evolution to determine if they were SN-like. If an object was deemed to be SN-like, we then estimated the date of peak brightness using the same polynomial fitting method previously described. As with the classified SNe, we use the date of peak brightness to determine if the SN-like object could be a YSN, a pre-peak SN, or a post-peak SN. 
Examples of non-SN contaminants in our YSN candidate samples are presented in Figure \ref{fig:ztf_unclass_contam_lcs}. Approximately $4$ percent of the contaminants are likely unclassified AGN or quasars (see left plot of Figure \ref{fig:ztf_unclass_contam_lcs}). For $79$ percent of contaminants, their nature is unclear due to the lack of observations, with many light curves having fewer than 5 observations. The remaining $17$ percent of contaminants have reasonably sampled light curves but an unclear nature (see middle and right plots of Figure \ref{fig:ztf_unclass_contam_lcs}).

Using the TNS classifications, the estimated dates of peak brightness for the SNe, and the results from our visual inspection of the unclassified objects, we calculated the contamination rate. As the aim of our YSN selection criteria is to select YSN, we considered all non-YSN objects to be contamination. Therefore, we calculated the contamination of a given sample as the percentage of non-YSN objects in the sample relative to the total number of objects in the sample.
Additionally, we calculated and provide the contamination rate where only non-SN objects (TDE, novae, and the unclassified objects that were not deemed SN-like) are considered contaminants.

\subsubsection{ZTF YSN Candidate Samples}\label{sec:ztf_samples_results}

\begin{table*}
    \setlength{\tabcolsep}{4pt}

    \caption{Six samples of objects that passed the selection criteria stated in Table \ref{tab:ztf_selection_criteria}, applying the different combinations (as stated) of age and brightening rate criterion thresholds. Provided are the TNS classifications of the objects, with the SNe being further classified based on if they were young, pre-peak, or post-peak at the time of selection. The unclassified objects were visually inspected, with the SN-like objects being further identified as being YSN-like, pre-peak SN-like, or post-peak SN-like. YSNe and YSN-like objects are a subset of pre-peak SNe and pre-peak SN-like objects respectively.
    Also provided are the contamination rates of the samples, with one assuming non-YSNe as contaminants and the other assuming non-SNe as contaminants.
    \\ * Classifications based on visual inspection of the objects' light curves}
    \label{tab:ztf_samples}
    \centering
    \begin{tabular}{@{\hskip 3pt \vline \hskip 3pt} c | c @{\hskip 3pt \vline \hskip 3pt} c @{\hskip 3pt \vline \hskip 3pt} c @{\hskip 3pt \vline \hskip 3pt} c @{\hskip 3pt \vline \hskip 3pt} c @{\hskip 3pt \vline \hskip 3pt}  c @{\hskip 3pt \vline \hskip 3pt} c | c | c | c @{\hskip 3pt \vline \hskip 3pt} c | c @{\hskip 3pt \vline \hskip 3pt}}
        \hline
        \hline
            \multicolumn{2}{@{\hskip 3pt \vline \hskip 3pt} c @{\hskip 3pt \vline \hskip 3pt}}{Criteria Thresholds} & \multicolumn{9}{c @{\hskip 3pt \vline \hskip 3pt}}{Classification} & \multicolumn{2}{c @{\hskip 3pt \vline \hskip 3pt}}{Contamination} \\
             
            Age & Brightening Rate & YSNe & Pre-Peak & Post-Peak & TDE & Nova & \multicolumn{4}{c @{\hskip 3pt \vline \hskip 3pt}}{Unclassified} & Non-YSNe & Non-SNe \\

            & & & SNe & SNe & & & Non-SNe* & YSNe* & Pre-Peak SNe* & Post-Peak SNe* & & \\
        \hline
        \hline
            $<7$ days & $> 0.05$ mag/day & 79 & 179 & 10 & 2 & 1 & 308 & 51 & 443 & 36 & 87$\%$ & 32$\%$ \\
            $<14$ days & $> 0.05$ mag/day & 84 & 212 & 17 & 3 & 1 & 407 & 57 & 531 & 87 & 89$\%$ & 33$\%$  \\
            $<7$ days & $> 0.1$ mag/day & 74 & 149 & 2 & 1 & 1 & 172 & 37 & 307 & 14 & 83$\%$ & 27$\%$ \\
            $<14$ days & $> 0.1$ mag/day & 76 & 166 & 5 & 2 & 1 & 216 & 40 & 347 & 32 & 85$\%$ & 28$\%$ \\
            $<7$ days & $> 0.2$ mag/day & 60 & 97 & 1 & 1 & 1 & 58 & 17 & 100 & 6 & 71$\%$ & 23$\%$\\
            $<14$ days & $> 0.2$ mag/day & 60 & 99 & 1 & 2 & 1 & 69 & 17 & 104 & 11 & 73$\%$ & 25$\%$\\
        \hline
        \hline 
        
    \end{tabular}
\end{table*}

Presented in Table \ref{tab:ztf_samples} are the six YSN candidate samples produced from the proposed selection criteria with different combinations of the age and brightening rate criteria threshold values. The table displays the number of objects based on their TNS classifications, with the SNe being further determined to be YSNe, pre-peak SNe, or post-peak SNe. The number of unclassified objects is broken down further into non-SNe, YSNe, pre-peak SNe, or post-peak SNe, which is based on our visual inspection of the unclassified objects' light curves.
Additionally, the table also provides the non-YSN contamination rates (percentage of the sample that is not a YSN) and non-SN contamination rates (percentage of the sample that is not a SN) of the samples. The non-YSN and non-SN contamination rates have estimated uncertainties on the order of $\pm5$ percent and $\pm2$ percent respectively.

As can be seen, Table \ref{tab:ztf_samples} shows that the sample produced from the selection criteria with an age $<14$ days and brightening rate $>0.05$ mag/day produces the largest YSN candidate sample. It contains 84 TNS classified YSNe (SNe selected before a phase of $-10$ days), with a further 57 YSNe from visual inspection of the unclassified objects' light curves. However, this sample is also the most contaminated, with non-YSN and non-SN contamination rates of 89 percent and 33 percent respectively.
In contrast, the smallest but least contaminated sample is that produced using the age and brightening rate criteria thresholds of 7 days and 0.2 mag/day respectively. This sample contains 60 classified YSN and 17 visually identified YSNe, and has non-YSN and non-SN contamination rates of 71 percent and 23 percent respectively. 

More generally, Table \ref{tab:ztf_samples} shows that the samples produced using the age constraint of 14 days contain more YSNe and SNe than those produced from the 7 day constraint. However, the samples produced using the 14 day constraint are up to 2 percent more contaminated, in terms of both non-YSN and non-SN contamination, than the corresponding samples produced with the 7 day constraint.
Furthermore, the results in Table \ref{tab:ztf_samples} show that as the brightening rate threshold value increases, the contamination and number of YSNe/SNe selected decreases.

\subsubsection{Evaluation of Selection Criteria}\label{subsubsec:ztf_evalutation}

From our results presented in Section \ref{sec:ztf_samples_results}, it is clear that the purest sample is that produced using the age and brightening rate thresholds of 7 days and 0.2 mag/day respectively. As the aim of our selection criteria is to produce a pure sample of YSN, we select these thresholds for our YSN selection criteria. We will now discuss in more detail the YSN candidate sample produced using the age and brightening rate thresholds of 7 days and 0.2 mag/day respectively. For simplicity, hereafter we refer to this sample and the selected thresholds as the YSN candidate sample and the YSN selection criteria respectively.

Although our YSN candidate sample has the lowest contamination of the samples produced, it is still very contaminated with a non-YSN contamination of 71 percent. The majority of the non-YSN contaminants are SNe that were selected after a phase of $-10$ days, accounting for 127 of the 187 contaminants (67.91 percent). This is reflected in the non-SN contamination rate, which is only 23 percent.

At a first glance a YSN candidate sample with a non-SN contamination rate of 23 percent might seem to be too high to be worth implementing into a TiDES-like survey. However, this percentage does not quantify the number of fibre hours that would be spent on the YSN candidate sample contaminants. By calculating the wasted fibre hours as a percentage of TiDES's total fibre hours we can evaluate if this contamination level is too high or not.
While we can calculate the number of fibre hours that would be wasted based on our ZTF YSN candidate sample, this would not be representative of the actual number that would be wasted during the operation of a TiDES-like survey. This is because TiDES-like surveys, and hence our selection criteria, will be selecting objects from LSST as opposed to ZTF. LSST will produce a much larger YSN candidate sample than that produced from the ZTF alerts. Therefore, we revisit this discussion of wasted fibre hours in Section \ref{subsec:lsst_discussion_contamination}, where we used the LSST simulations to produce a YSN candidate sample.   

\begin{figure}
    \centering
    \includegraphics[width=\linewidth]{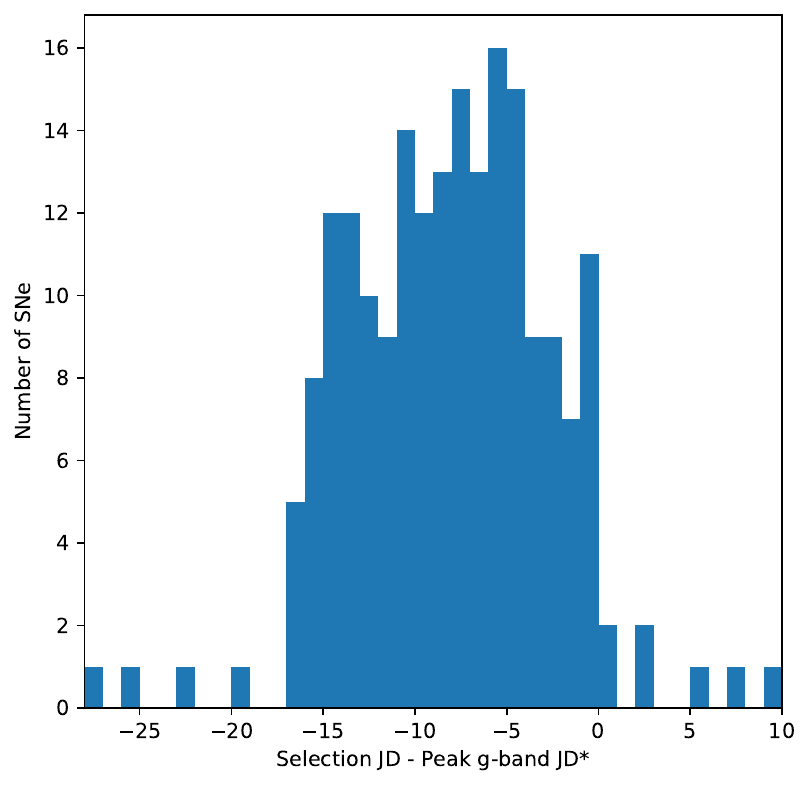}
    \caption{Phase distribution of the TNS classified and visually determined SNe in the sample that was produced by applying our YSN selection criteria (see Table \ref{tab:ztf_selection_criteria}), with age and brightening rate thresholds of 7 days and 0.2 mag/day respectively, to 60 nights of ZTF transient alerts.
    Note that the phase has been truncated at -28 days.}
    \label{fig:ztf_phase_distro}
\end{figure}

In addition to investigating the contamination of our YSN candidate sample, we also looked at the phase at which the SNe (both TNS classified SNe and SNe identified from our visual inspection) were selected by our selection criteria. Presented in Figure \ref{fig:ztf_phase_distro} is the selected phase distribution of the SNe in our YSN candidate sample. It should be noted that the presented phases are not definitive phases, as they were calculated using estimates of the peak JDs of the SNe. As can be seen from Figure \ref{fig:ztf_phase_distro}, almost all of the SNe, with the exception of 7, were selected at pre-peak phases.
However, only 77 of the 204 SNe (37.75 percent) are YSNe, or in other words were selected before a phase of $-10$ days. This might seem like a low percentage of YSN, however, to fully evaluate our YSN selection criteria's performance we need to investigate how they perform when applied to LSST, which we present in Section \ref{sec:lsst_application}.

\section{Application of YSN Selection Criteria to LSST}\label{sec:lsst_application}

To investigate the impact that our YSN selection criteria could have on a TiDES-like transient programme, we can not use the ZTF alerts as they are not fully representative of the future LSST alerts that TiDES-like surveys will draw their targets from. Some of the differences between the two surveys that will affect our selection criteria and the resulting objects are as follows: LSST has different filter bands ($ugrizY$- bands), LSST has a much fainter limiting magnitude than ZTF, LSST will have a different observing strategy to that of ZTF, and LSST will observe SNe at a higher redshift distribution than ZTF. With this in mind, we adapted the ZTF developed YSN selection criteria for use with LSST data, evaluating their performance by applying them to an LSST simulation. Although there are many different LSST simulations that are suitable for use in this study, for example \texttt{PLAsTiCC} \citep{Kessler_2019} or \texttt{ELAsTiCC} \footnote{The DESC ELAsTiCC Challenge \url{https://portal.nersc.gov/cfs/lsst/DESC_TD_PUBLIC/ELASTICC/} (accessed on 10/07/2025)}, we chose to use the simulations of \citetalias{TiDES_2025}. This was to allow for a direct comparison between our YSN selection criteria and those of TiDES (\citetalias{TiDES_2025}).

\subsection{Producing an LSST YSN Candidate Sample}\label{subsec:lsst_methods}

\subsubsection{Adaptation of ZTF Developed YSN Selection Criteria}

With the addition of the $i$-, $z$-, $u$-, and $Y$-bands of LSST, we adapted the ZTF developed YSN selection criteria by including the LSST $i$- and $z$-bands. We did not include the $u$-band in our selection criteria as it is not often used by LSST and so our selection criteria will not benefit from including it. As we utilised the LSST simulation of \citetalias{TiDES_2025}, the $Y$-band was also not included. This is due to the rest-frame $Y$-band being poorly defined in the spectrophotometric models that were used in this LSST simulation.
To incorporate the $i$- and $z$-bands in our selection criteria, we included them in our applied magnitude criterion, as well as adapting our age and brightening rate criteria so that observations in the $i$- and $z$-bands are also considered if available.

One significant difference between ZTF and LSST will be the redshift ($z$) distribution of the observed transients. SNe in the ZTF sample typically have redshifts below 0.1, although there are some with redshifts up to 0.3 \citep{Rigault_2025}. In contrast, the
LSST SN redshift distribution should greatly exceed that of ZTF. For example, the LSST SN Ia distribution is estimated to peak at $z{\sim}0.5$, with some detected at redshifts beyond $z=1$ \citep{Kessler_2019, Petrecca_2024}. 
Due to redshift distribution difference between ZTF and LSST, the brightening rate should be altered accordingly. For example, this could be achieved by applying the redshift correction given by equation \ref{eqn:br_redshift_correction}.
\begin{equation}
    \label{eqn:br_redshift_correction}
    \rm{BR}_{\rm{LSST}} = \rm{BR}_{\rm{ZTF}} / (1 + z_{\rm{LSST}})
\end{equation}
where $\rm{BR}$ is the brightening rate applied to either LSST or ZTF ($\rm{BR}_{\rm{ZTF}}=0.2$), and $z_{\rm{LSST}}$ is the average redshift of the LSST redshift distribution.
An issue with adapting the brightening rate threshold based on the LSST SN redshift distribution is that it will be decreased. For example, if we assume $z_{\rm{LSST}} = 0.5$ then $\rm{BR}_{\rm{LSST}}$ is 0.13. As we showed in Section \ref{sec:ztf_samples_results}, decreasing the brightening rate increases the contamination of the sample. Therefore, in the interest of producing a pure YSN sample, we do not adapt the brightening rate. This has the implication that our LSST selected YSN candidate sample will have a lower redshift distribution than if we were to adapt the brightening rate criterion. 

Additionally, we applied a declination constraint of $-70$\degree $<$ Dec $<5$\degree, so that only targets within the 4MOST footprint are selected from the LSST surveys.
The adapted YSN selection criteria for LSST are stated, with reasoning, in Table \ref{tab:lsst_selection_criteria}. It is worth noting that the Sherlock classification criterion was not applied as it is only available through the Lasair broker. Regardless, we state this criterion in Table \ref{tab:lsst_selection_criteria} as it should be applied to the future LSST live transient alerts to help exclude contamination from the produced YSN candidate sample.

\begin{table*}
    \caption{
        Selection criteria that were applied to the LSST simulations of transient events to select YSNe. Provided are the criteria along with a description of their purpose. 
        \\ * This criterion is not applied to the LSST simulations but will be applied to LSST live alerts in the future.}
    \label{tab:lsst_selection_criteria}
    \centering
    \begin{tabular}{c|c}
        \hline
        \hline
            Criterion & Reasoning \\
        \hline
        \hline
            $g$-, $r$-, $i$-, or $z$-band magnitude $<22.5$ mag & Ensures that the object will meet the TiDES SN SSC \\
            $-70$\degree$<$ Declination $<5$\degree & Extent of 4MOST declination range \\
            Galactic latitude $<-10$\degree OR Galactic latitude $>10$\degree & Removal of Galactic transient sources \\
            Number of $g$-, $r$-, $i$-, or $z$-band $>5 \sigma$ positive difference detections $\geq 2$ & Two or more $>5 \sigma$ detections in a band required for brightening rate criterion \\
            \texttt{Sherlock} classification not: VS, AGN, CV, BS * & Removal of contaminants \\
            Age < 14 days & Removes older objects that are unlikely pre-peak SNe \\
            Brightening rate > $0.2$ mag/day & Removes dimming and slow brightening sources that are unlikely to be YSNe. \\      
        \hline
        \hline 
    \end{tabular}
\end{table*}

\subsubsection{Applying the Selection Criteria}\label{subsubsec:applying_lsst_criteria}

With the criteria adapted for use with LSST, we applied them to the LSST simulation used by \citetalias{TiDES_2025}. This simulation utilises the LSST baseline V3.4 simulation\footnote{\url{https://survey-strategy.lsst.io/index.html} (accessed on 05/12/2024)} and the {\sl SuperNova ANAlysis} software \citep[SNANA;][]{kessler_2009} to produce a catalogue of SN events and photometry as would be observed by LSST over a 5 year period. The transients included in this simulation are SNe Ia, SNe Iax, SNe 91bg, SNe Ib, SNe Ic, SNe Ic-BL, SNe II, SNe IIb, SNe IIn, superluminous SNe (SLSNe), calcium rich transients (CART), and TDE. The simulations only include the LSST WFD and DDF surveys. The WFD survey covers an 18000 deg$^2$ area that, under currently plans, will be observed using a rolling cadence\footnote{LSST Survey Cadence Optimization Committee’s Phase 2 Recommendations, available at \url{https://pstn-055.lsst.io/} (accessed on 28/04/2025)} in $ugrizY$ filter bands down to a depth of 25 mag in the $g-$band \citep{lsst_reqs}. The DDF survey is a much smaller area survey consisting of five circular fields\footnote{Information of the 5th field: \url{https://community.lsst.org/t/scoc-endorsement-of-euclid-deep-field-south-observations/6406} (accessed on 25/09/2024)} of diameter ${\sim}3.5$\degree that will have a higher cadence and a deeper coverage than that of the WFD survey \citep{lsst_reqs}.

\subsection{LSST YSN Candidate Samples}\label{subsec:lsst_results}

\subsubsection{LSST WFD Survey} \label{subsubsec:lsst_wfd_results}

\begin{table*}
        \caption{Resulting number of transients selected from the 5 year LSST WFD survey simulation by our YSN selection criteria (stated in Table \ref{tab:lsst_selection_criteria}) and the \citetalias{TiDES_2025} selection criteria. Additionally, provided (in brackets) is the percentage of transients that were selected from the total number of LSST WFD survey simulated transients. Note that young transients are a subset of pre-peak transients.}
    \label{tab:wfd_samples}
    \centering
    \begin{tabular}{@{\hskip 5pt \vline \hskip 5pt} c @{\hskip 5pt \vline \hskip 5pt} r @{\hskip 5pt \vline \hskip 5pt} r | r | r @{\hskip 5pt \vline \hskip 5pt} r | r | r @{\hskip 5pt \vline \hskip 5pt}}
        \hline
        \hline 
            \multicolumn{8}{@{\hskip 5pt \vline \hskip 5pt} c @{\hskip 5pt \vline \hskip 5pt}}{5 Year WFD Survey} \\
        \hline
        \hline            
            Classification & Total Simulated & \multicolumn{3}{c @{\hskip 5pt \vline \hskip 5pt}}{YSN Candidate Sample} & \multicolumn{3}{c @{\hskip 5pt \vline \hskip 5pt}}{\citetalias{TiDES_2025} Sample} \\
            
            & & \multicolumn{1}{c}{Young} & \multicolumn{1}{c}{Pre-Peak} & \multicolumn{1}{c @{\hskip 5pt \vline \hskip 5pt}}{Post-Peak} & \multicolumn{1}{c}{Young} & \multicolumn{1}{c}{Pre-Peak} & \multicolumn{1}{c @{\hskip 5pt \vline \hskip 5pt}}{Post-Peak} \\
            
            & & \multicolumn{1}{c}{($<-10$ days)} & \multicolumn{1}{c}{($<0$ days)} & \multicolumn{1}{c @{\hskip 5pt \vline \hskip 5pt}}{($\geq0$ days)} & \multicolumn{1}{c}{($<-10$ days)} & \multicolumn{1}{c}{($<0$ days)} & \multicolumn{1}{c @{\hskip 5pt \vline \hskip 5pt}}{($\geq0$ days)} \\
        \hline
        \hline
            Ia & 2971223 & 36168 (1.22\%) & 39533 (1.33\%) & 0 (0.00\%) & 67376 (2.27\%) & 316269 (10.64\%) & 199698 (6.72\%) \\
            Iax & 83351 & 529 (0.63\%) & 614 (0.74\%) & 0 (0.00\%) & 1514 (1.82\%) & 5815 (7.00\%) & 5105 (6.12\%) \\
            91bg & 69266 & 513 (0.74\%) & 2649 (3.82\%) & 0 (0.00\%) & 380 (0.55\%)& 9664 (13.95\%) & 10956 (15.82\%) \\
            Ib & 89365 & 515 (0.58\%) & 675 (0.76\%) & 0 (0.00\%) & 2641 (2.96\%) & 8095 (9.06\%) & 7303 (8.17\%) \\
            Ic & 53670 & 196 (0.37\%) & 710 (1.32\%) & 0 (0.00\%) & 751 (1.40\%) & 3734 (6.96\%) & 5379 (10.02\%) \\
            Ic-BL & 34511 & 141 (0.41\%) & 344 (1.00\%) & 51 (0.15\%) & 217 (0.63\%) & 1968 (5.70\%) & 3896 (11.29\%) \\
            II & 782637 & 4 (0.00\%) & 2939 (0.38\%) & 774 (0.10\%) & 19 (0.00\%) & 4741 (0.61\%) & 74885 (9.57\%) \\
            IIb & 220409 & 1218 (0.55\%) & 3891 (1.77\%) & 962 (0.44\%) & 2465 (1.12\%) & 13594 (6.17\%) & 28744 (13.04\%) \\
            IIn & 483957 & 60 (0.01\%) & 2561 (0.53\%) & 105 (0.02\%) & 3352 (0.69\%) & 11728 (2.42\%) & 33555 (6.93\%)\\
            SLSN & 32922 & 285 (0.87\%) & 301 (0.91\%) & 0 (0.00\%) & 7787 (23.65\%) & 9709 (29.49\%) & 8600 (26.12\%) \\
            CART & 16719 & 30 (0.18\%) & 141 (0.84\%) & 10 (0.06\%) & 96 (0.57\%) & 589 (3.52\%) & 1640 (9.81\%) \\
            TDE & 23476 & 148 (0.63\%) & 148 (0.63\%) & 0 (0.00\%) & 1409 (6.00\%) & 2261 (9.63\%) & 1523 (6.49\%) \\
        \hline
            Total & 4861506 & 39807 (0.82\%) & 54506 (1.12\%) & 1902 (0.04\%) & 88007 (1.81\%) & 388167 (7.98\%) & 381284 (7.84\%) \\
        \hline
        \hline 
    \end{tabular}
\end{table*}

Following the methods in Section \ref{subsec:lsst_methods}, presented in Table \ref{tab:wfd_samples} are the resulting samples of selected transients covering 5 years of the LSST WFD survey. This shows the total number of objects simulated, the number of objects selected by our YSN selection criteria (YSN candidate sample). For comparison, we present the number of objects selected by the current TiDES selection criteria used by \citetalias{TiDES_2025}, hereafter referred to as the \citetalias{TiDES_2025} selection criteria and \citetalias{TiDES_2025} sample. We have also defined whether the objects were young (phase $<-10$ days), pre-peak (phase $<0$ days), or post-peak (phase $\geq0$ days) at the time of selection. Note that YSNe are a subset of pre-peak SNe.
The phase of selection was calculated using the peak date provided in the LSST simulation.
As can be seen from Table \ref{tab:wfd_samples}, our YSN selection criteria selected $56408$ transients, of which $39807$ were young at the time of selection. The YSN candidate sample size is approximately only $7$ percent of the size of the current simulated \citetalias{TiDES_2025} sample, and contains $1$ percent of the total number of simulated WFD survey transients.

\begin{figure}
    \centering
    \includegraphics[width=\linewidth]{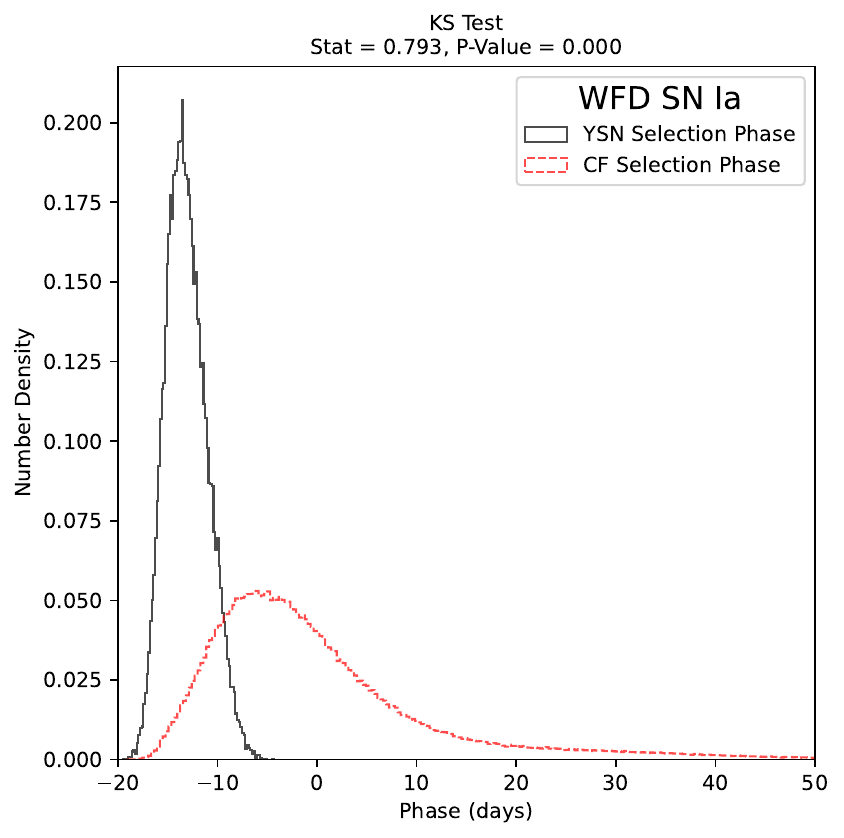}
    \caption{Comparison between the SN Ia selection phase distributions produced by applying our selection criteria (YSN; see Table \ref{tab:lsst_selection_criteria}) and the \citetalias{TiDES_2025} selection criteria to the LSST WFD survey simulation. Note that the distributions are normalised and that the phase has been truncated at 50 days.}
    \label{fig:wfd_ia_phase_distros}
\end{figure}

\begin{figure}
    \centering
    \includegraphics[width=\linewidth]{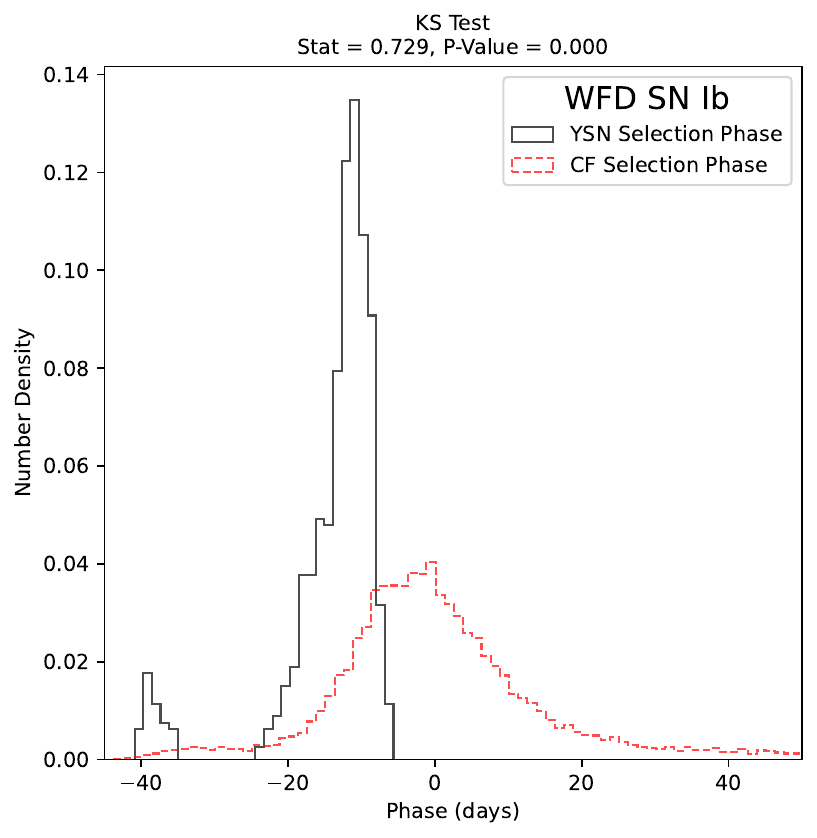}
    \caption{Comparison between the SN Ib selection phase distributions produced by applying our selection criteria (YSN; see Table \ref{tab:lsst_selection_criteria}) and the \citetalias{TiDES_2025} selection criteria to the LSST WFD survey simulation. Note that the distributions are normalised and that the phase has been truncated at 50 days.}
    \label{fig:wfd_ib_phase_distros}
\end{figure}

\begin{figure}
    \centering
    \includegraphics[width=\linewidth]{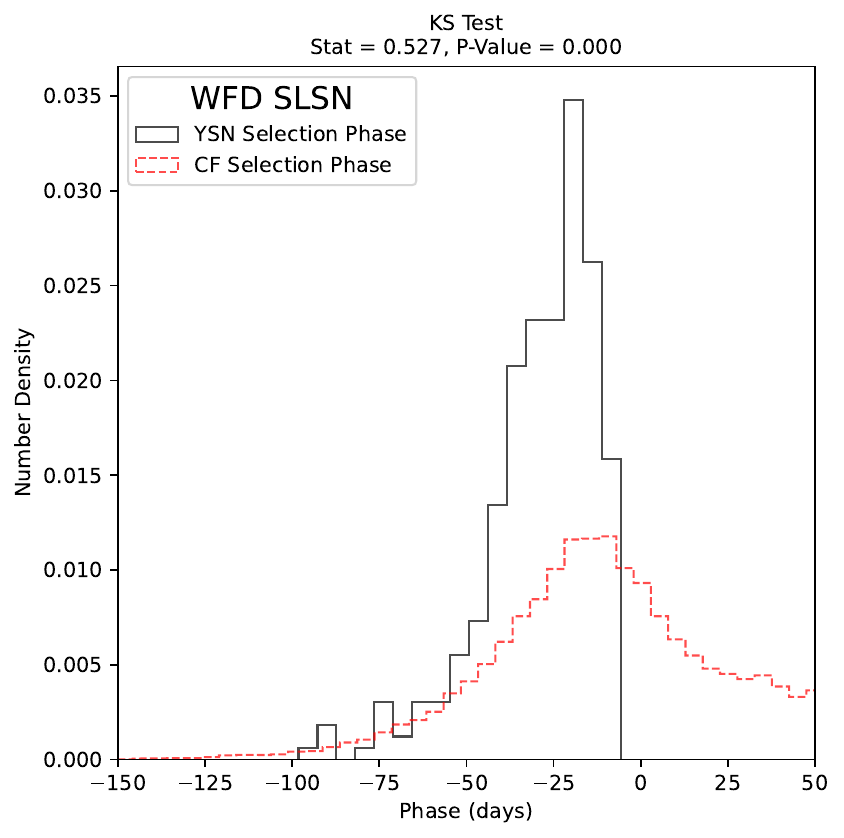}
    \caption{Comparison between the SLSN selection phase distributions produced by applying our selection criteria (YSN; see Table \ref{tab:lsst_selection_criteria}) and the \citetalias{TiDES_2025} selection criteria to the LSST WFD survey simulation. Note that the distributions are normalised and that the phase has been truncated at -150 days and 50 days.}
    \label{fig:wfd_slsn_phase_distros}
\end{figure}

\begin{figure*}
    \begin{tabular}{cc}
         \includegraphics[width=0.47\linewidth]{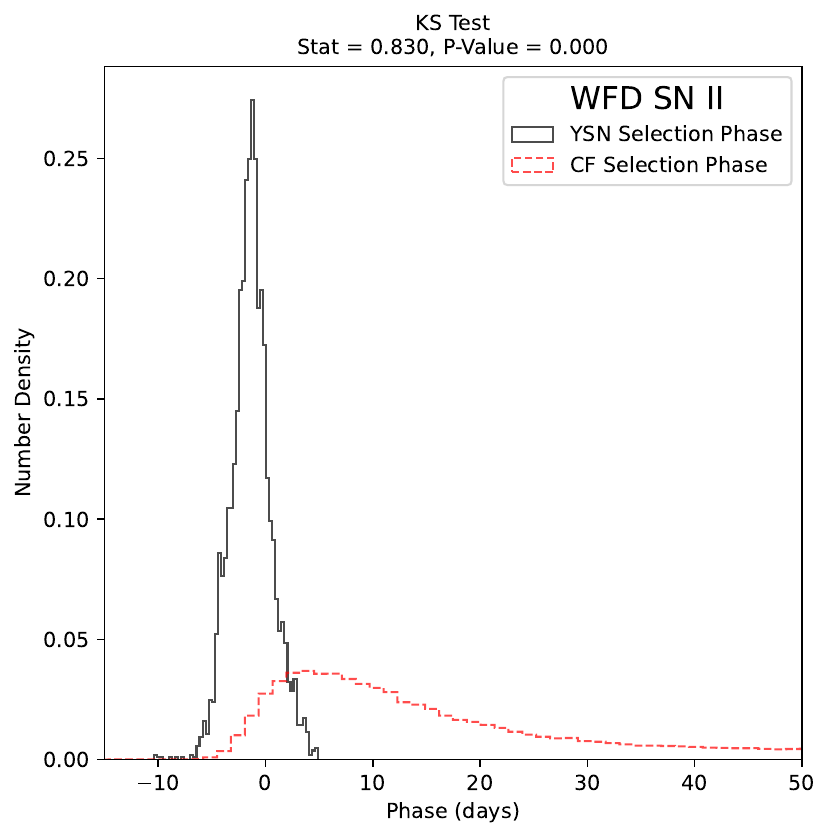} & \includegraphics[width=0.47\linewidth]{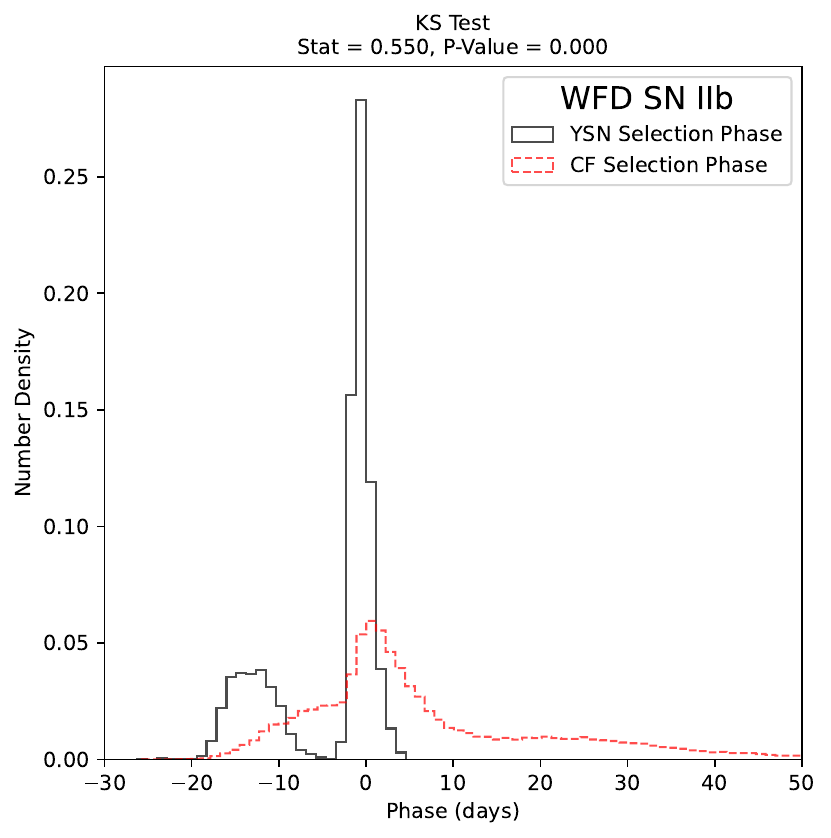} \\\includegraphics[width=0.47\linewidth]{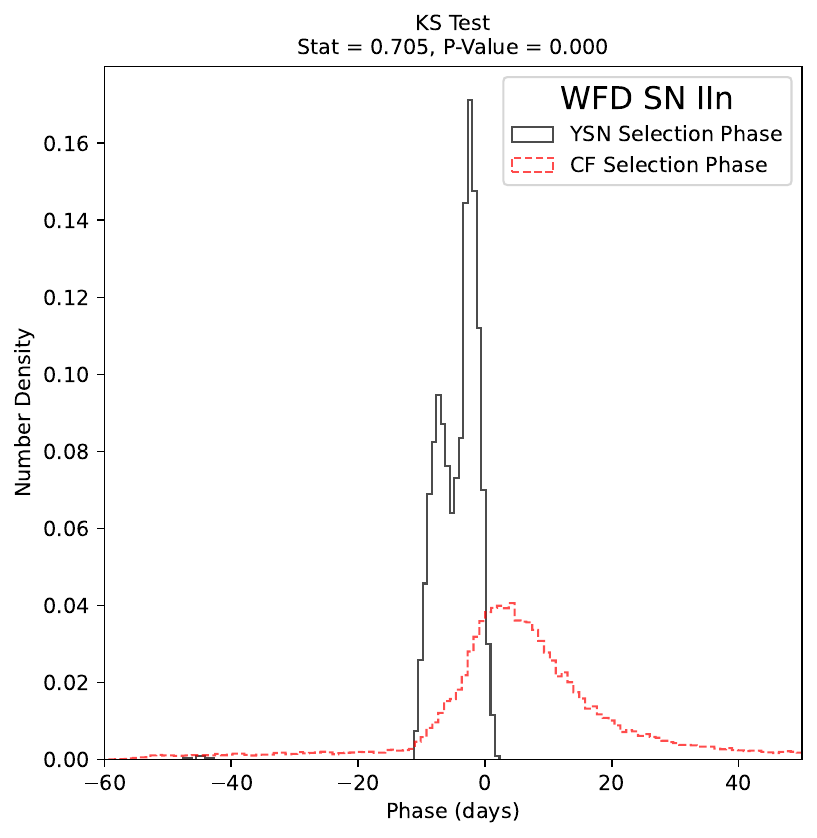} & \includegraphics[width=0.47\linewidth]{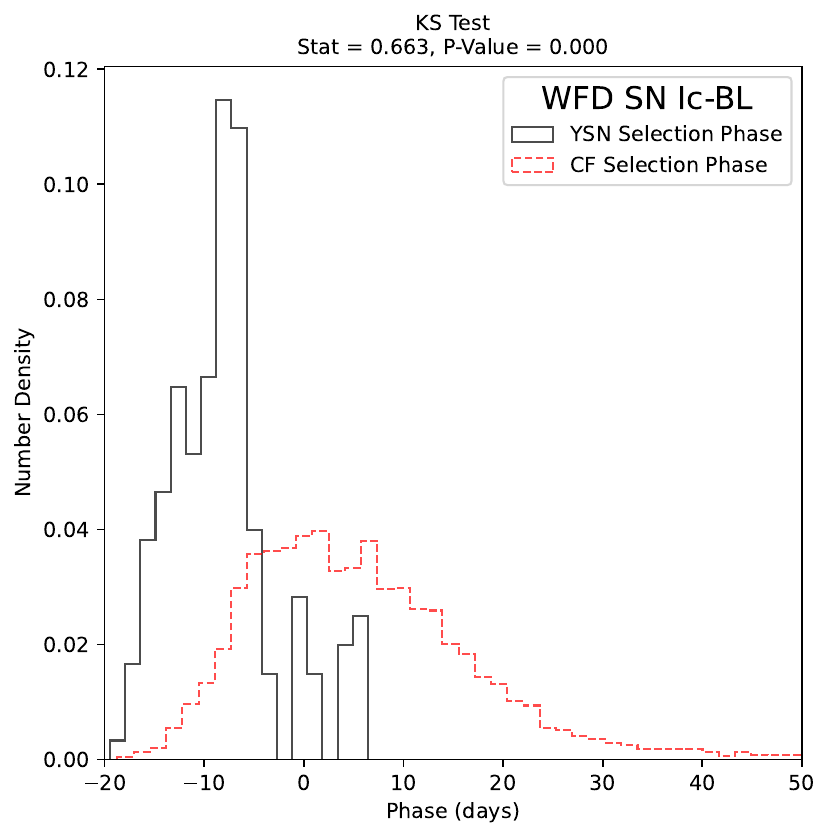}
    \end{tabular}
    
    \caption{Comparison between the SN II, IIb, IIn, and IC-BL selection phase distributions produced by applying our selection criteria (YSN; see Table \ref{tab:lsst_selection_criteria}) and the \citetalias{TiDES_2025} selection criteria to the LSST WFD survey simulation. Note that the distributions are normalised and that all phases have been truncated at 50 days.}
    \label{fig:wfd_ii_iib_iin_ic-bl_phase_distros}
\end{figure*}

For each classification of transient within our sample, we present in Figures \ref{fig:wfd_ia_phase_distros}-\ref{fig:wfd_ii_iib_iin_ic-bl_phase_distros} and Appendix \ref{app:wfd_phase_distros} comparisons between the selection phases for the samples produced by our selection criteria and the \citetalias{TiDES_2025} selection criteria. In general, the peak of the selection phase distributions produced from our selection criteria occur a few days before those produced by the \citetalias{TiDES_2025} selection criteria. For example, Figure \ref{fig:wfd_ia_phase_distros} shows that the SN Ia selection phase distribution produced by our YSN criteria peaks at ${\sim-14}$ days while the distribution produced by the \citetalias{TiDES_2025} peaks at ${\sim}-6$ days.
Furthermore, the two sided Kolmogorov–Smirnov (KS) tests performed on the YSN and \citetalias{TiDES_2025} phase distributions for each transient class all return P-values less than 0.0005, indicating that the two SN samples are drawn from different distributions.

For the majority of the transient classes, the selection phase distributions produced by our selection criteria are unimodal. However, for SNe Ib (Figure \ref{fig:wfd_ib_phase_distros}) and SNe IIb (Figure \ref{fig:wfd_ii_iib_iin_ic-bl_phase_distros}) their selection phase distributions display bimodality with the stronger of the peaks occurring ${\sim}25$ and ${\sim}15$ days after the weaker peaks respectively. The weaker of the peaks occur at a phase of ${\sim}-40$ days for SNe Ib and ${\sim}-15$ days for SNe IIb. The cause of the bimodality in the SN IIb selection phase distribution is unclear. However, for the SN Ib distribution it arises due to the inclusion of SN 2005bf in the spectrophotometric templates used to simulate the transients in the LSST simulation. SN 2005bf is a transitional SN (Ib to Ic) and has a unique morphology, exhibiting two maxima separated by about 25 days and an unusually long rise to peak brightness of ${\sim}40$ days \citep{Folatelli_2006}.

Lastly, the selection phase distributions produced by our selection criteria for many of the transient classes do not exceed into positive phases (post-peak), as is seen in Figures \ref{fig:wfd_ia_phase_distros}, \ref{fig:wfd_ib_phase_distros}, and \ref{fig:wfd_slsn_phase_distros}. However, for SNe II, IIb, IIn, and Ic-BL (see Figure \ref{fig:wfd_ii_iib_iin_ic-bl_phase_distros}) the selection phase distributions extend up to phases of ${\sim} 7$ days.

To gain a better understanding of the crossover between the objects selected by our YSN criteria and the \citetalias{TiDES_2025} criteria, we present in Table \ref{tab:wfd_ysne_tides_objects_numbers} the number of selected transients that are common to both sets of selection criteria. We also include the number of transients that are only selected by our selection criteria.
As is seen, there are $38657$ SNe Ia and $15787$ non-SNe Ia that are commonly selected by both sets of selection criteria. Also indicated is that our selection criteria selects an additional $876$ SNe Ia and $1088$ non-SNe Ia that the \citetalias{TiDES_2025} selection criteria does not select.

\begin{table}
        \caption{Number of LSST WFD simulated transients that were commonly selected by both our YSN selection criteria (see Table \ref{tab:lsst_selection_criteria}) and the \citetalias{TiDES_2025} selection criteria. Also included is the number of simulated transients that were only selected by our YSN criteria. Additionally, provided for the commonly selected transients are the average phases of selection by the \citetalias{TiDES_2025} criteria and our YSN criteria.}
    \label{tab:wfd_ysne_tides_objects_numbers}
    \centering
    \begin{tabular}{@{\hskip 3pt \vline \hskip 3pt} 
                c @{\hskip 3pt \vline \hskip 3pt} 
                c @{\hskip 3pt \vline \hskip 3pt}
                r |
                r @{\hskip 10pt \vline \hskip 3pt} 
                c @{\hskip 3pt \vline \hskip 3pt}}
    \hline
    \hline
    Classification 
        & \multicolumn{3}{c @{\hskip 3pt \vline \hskip 3pt}}{Commonly Selected} 
        & \multicolumn{1}{c @{\hskip 3pt \vline \hskip 3pt}}{YSN Uniquely} \\
    
        & \multicolumn{1}{c @{\hskip 3pt \vline \hskip 3pt}}{Quantity} 
        & \multicolumn{2}{c @{\hskip 3pt \vline \hskip 3pt}}{Mean Selection Phase} 
        & \multicolumn{1}{c @{\hskip 3pt \vline \hskip 3pt}}{Selected} \\

        &
        & \multicolumn{1}{c @{\hskip 10pt}}{\citetalias{TiDES_2025}} 
        & \multicolumn{1}{c @{\hskip 3pt \vline \hskip 3pt}}{YSN} 
        & \multicolumn{1}{c @{\hskip 3pt \vline \hskip 3pt}}{} \\
    
    \hline
    \hline
            Ia & 38657 & $-9$ d &  $-13$ d & 876 \\
            Iax & 604 & $-10$ d & $-14$ d & 10 \\
            91bg & 2540 & $-5$ d & $-8$ d & 109 \\
            Ib & 661 & $-10$ d & $-14$ d & 14  \\
            Ic & 694 & $-6$ d & $-9$ d & 16  \\
            Ic-BL & 383 & $-3$ d & $-8$ d & 12  \\
            II & 3637 & $3$ d & $-1$ d & 76  \\
            IIb& 4078 & $-1$ d & $-5$ d & 775  \\
            IIn & 2598 & $-1$ d & $-4$ d & 68  \\
            SLSN & 299 & $-21$ d & $-29$ d & 2  \\
            CART & 148 & $-2$ d & $-6$ d & 3  \\
            TDE & 145 & $-20$ d & $-25$ d & 3  \\
            \hline
            Total & 54444 & & & 1964 \\
        \hline
        \hline 
    \end{tabular}
\end{table}

\begin{figure}
    \centering
    \includegraphics[width=\linewidth]{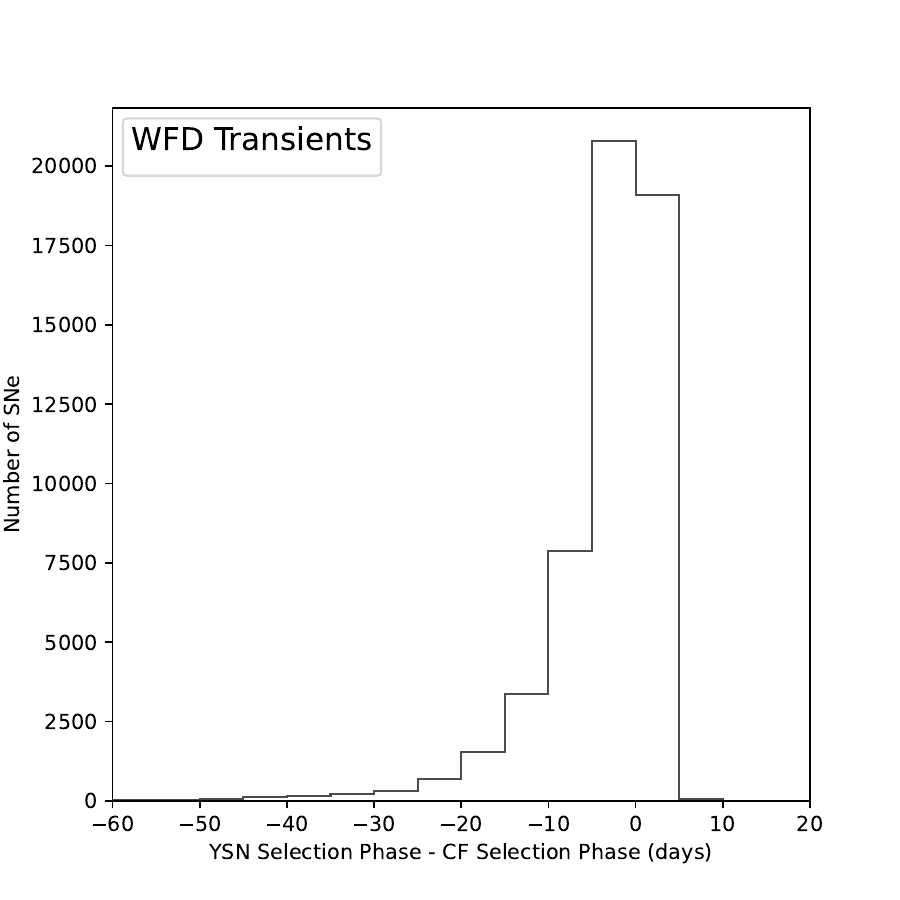}
    \caption{Distribution of the time differences of the selection of the LSST WFD transients that were selected by both our YSN selection criteria and the \citetalias{TiDES_2025} criteria. All of the simulated transient types (see Table \ref{tab:wfd_samples}) are included. Note that the phase difference has been truncated at -60 days.}
    \label{fig:wfd_common_trigger_diff}
\end{figure}

Additionally, for each of the commonly selected LSST WFD survey transients, we compared the difference between their selection phases when selected by our YSN selection criteria and the \citetalias{TiDES_2025} criteria. Presented in Figure \ref{fig:wfd_common_trigger_diff} is the distribution of the selection phase differences. As is shown, the majority of transients (${\sim}35000$) have a negative time difference, or in other words were selected by the YSN selection criteria before the \citetalias{TiDES_2025} selection criteria. As can be seen from Table \ref{tab:wfd_ysne_tides_objects_numbers}, the commonly selected WFD transients are on average selected three to five days earlier by our YSN selection criteria.

\subsubsection{LSST DDF Survey}\label{subsubsec:lsst_ddf_results}

\begin{table*}
       \caption{Resulting number of transients selected from the 5 year LSST DDF survey simulation by our YSN selection criteria (stated in Table \ref{tab:lsst_selection_criteria}) and the \citetalias{TiDES_2025} selection criteria. Additionally, provided (in brackets) is the percentage of transients that were selected from the total number of LSST DDF survey simulated transients.
       Note that young transients are a subset of pre-peak transients.}
    \label{tab:ddf_samples}
    \centering
    \begin{tabular}{@{\hskip 5pt \vline \hskip 5pt} c @{\hskip 5pt \vline \hskip 5pt} r @{\hskip 5pt \vline \hskip 5pt} r | r | r @{\hskip 5pt \vline \hskip 5pt} r | r | r @{\hskip 5pt \vline \hskip 5pt}}
        \hline
        \hline 
            \multicolumn{8}{@{\hskip 5pt \vline \hskip 5pt} c @{\hskip 5pt \vline \hskip 5pt}}{5 Year DDF Survey} \\
        \hline
        \hline
            
            Classification & Total Simulated & \multicolumn{3}{c @{\hskip 5pt \vline \hskip 5pt}}{YSN Candidate Sample} & \multicolumn{3}{c @{\hskip 5pt \vline \hskip 5pt}}{\citetalias{TiDES_2025} sample} \\
            
            & & \multicolumn{1}{c}{Young} & \multicolumn{1}{c}{Pre-Peak} & \multicolumn{1}{c @{\hskip 5pt \vline \hskip 5pt}}{Post-Peak} & \multicolumn{1}{c}{Young} & \multicolumn{1}{c}{Pre-Peak} & \multicolumn{1}{c @{\hskip 5pt \vline \hskip 5pt}}{Post-Peak} \\

            & & \multicolumn{1}{c}{($<-10$ days)} & \multicolumn{1}{c}{($<0$ days)} & \multicolumn{1}{c @{\hskip 5pt \vline \hskip 5pt}}{($\geq0$ days)} & \multicolumn{1}{c}{($<-10$ days)} & \multicolumn{1}{c}{($<0$ days)} & \multicolumn{1}{c @{\hskip 5pt \vline \hskip 5pt}}{($\geq0$ days)} \\
        \hline
        \hline
            Ia & 57504 & 504 (0.88\%) & 550 (0.96\%) & 0 (0.00\%) & 1223 (2.13\%) & 3956 (6.88\%) & 1352 (2.35\%) \\
            Iax & 2384 & 5 (0.21\%) & 6 (0.25\%)& 0 (0.00\%) & 21 (0.88\%) & 63 (2.64\%) & 48 (2.01\%) \\
            91bg & 1532 & 6 (0.39\%) & 31 (2.02\%) & 0 (0.00\%) & 7 (0.46\%) & 142 (9.27\%) & 84 (5.48\%) \\
            Ib & 2558 & 2 (0.08\%) & 5 (0.20\%) & 0 (0.00\%) & 27 (1.06\%) & 96 (3.75\%) & 43 (1.68\%) \\
            Ic & 1665 & 2 (0.12\%) & 9 (0.54\%) & 0 (0.00\%) & 9 (0.54\%) & 44 (2.64\%) & 55 (3.30\%) \\
            Ic-BL & 797 & 3 (0.38\%) & 8 (1.00\%) & 1 (0.13\%) & 4 (0.50\%) & 26 (3.26\%) & 26 (3.26\%) \\
            II & 27260 & 0 (0.00\%) & 47 (0.17\%) & 11 (0.04\%) & 0 (0.00\%) & 81 (0.30\%) & 667 (2.45\%) \\
            IIb & 5057 & 17 (0.34\%) & 53 (1.05\%) & 18 (0.36\%) & 33 (0.65\%) & 176 (3.48\%) & 285 (5.64\%) \\
            IIn & 9853 & 1 (0.01\%) & 47 (0.48\%) & 1 (0.01\%) & 34 (0.35\%) & 156 (1.58\%) & 279 (2.83\%) \\
            SLSN & 228 & 4 (1.75\%) & 4 (1.75\%) & 0 (0.00\%) & 60 (26.32\%) & 74 (32.46\%) & 58 (25.44\%) \\
            CART & 583 & 0 (0.00\%) & 1 (0.17\%) & 0 (0.00\%) & 0 (0.00\%) & 2 (0.34\%) & 13 (2.23\%) \\
            TDE & 293 & 3 (1.02\%) & 3 (1.02\%) & 0 (0.00\%) & 15 (5.12\%) & 19 (6.48\%) & 12 (4.10\%) \\
        \hline
            Total & 109714 & 547 (0.50\%) & 663 (0.60\%)  & 31 (0.03\%)  & 1433 (1.31\%) & 4835 (4.41\%) & 2922 (2.66\%) \\
        \hline
        \hline 
    \end{tabular}
\end{table*}

Following the methods presented in Section \ref{subsec:lsst_methods}, presented in Table \ref{tab:ddf_samples} are the resulting YSN candidate and \citetalias{TiDES_2025} samples of selected transients from the 5 year LSST DDF survey. As is shown, our YSN selection criteria selected $694$ transients, of which only $31$ were selected at post-peak phases and $547$ were selected before a phase of $-10$ days (young). In contrast, the \citetalias{TiDES_2025} sample contains $7757$ transients, of which $2922$ were selected after peak brightness and $1433$ were young at the time of selection.   

\begin{figure}
    \centering
    \includegraphics[width=\linewidth]{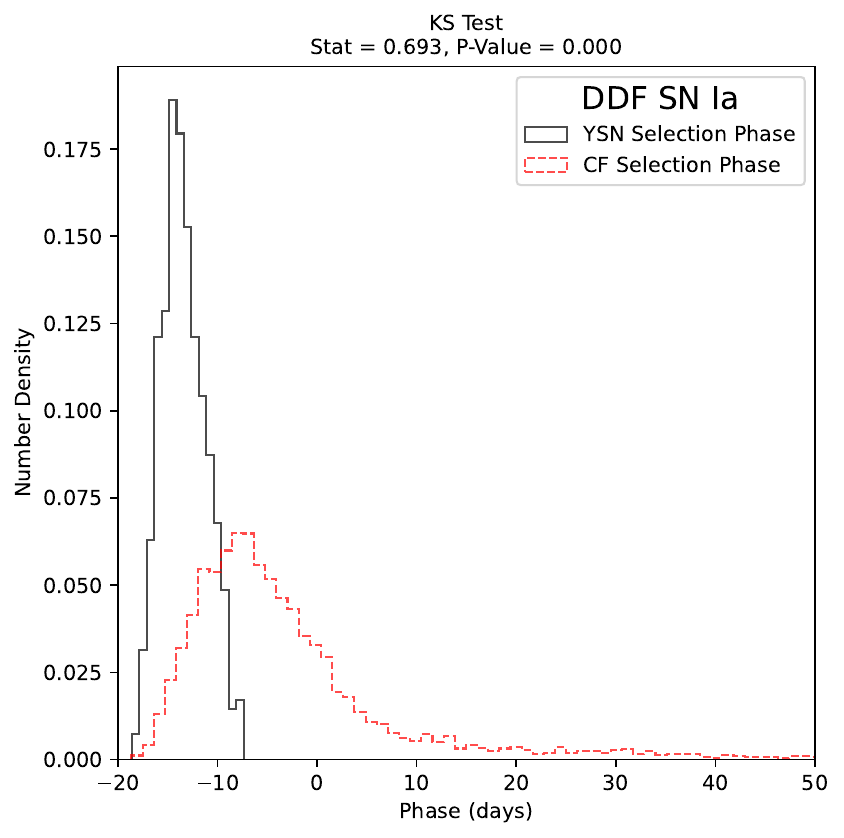}
    \caption{Comparison between the SN Ia selection phase distributions produced by applying our selection criteria (YSN; see Table \ref{tab:lsst_selection_criteria}) and the \citetalias{TiDES_2025} selection criteria to the LSST DDF survey simulation. Note that the distributions are normalised and that the phase has been truncated at 50 days.}
    \label{fig:ddf_ia_phase_distros}
\end{figure}

\begin{figure}
    \centering
    \includegraphics[width=\linewidth]{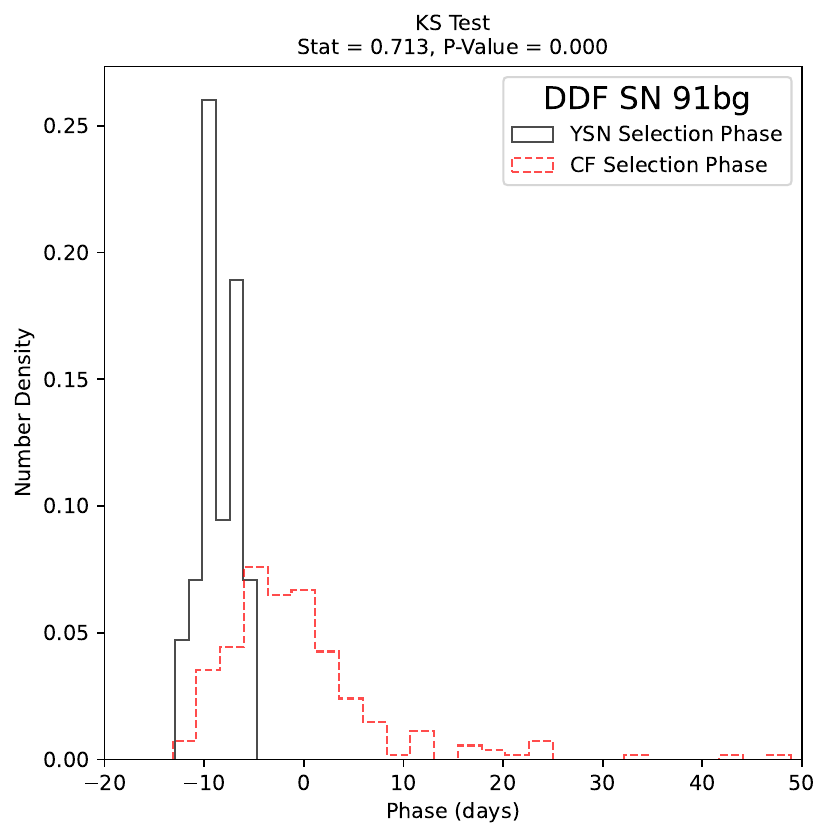}
    \caption{Comparison between the SN 91bg selection phase distributions produced by applying our selection criteria (YSN; see Table \ref{tab:lsst_selection_criteria}) and the \citetalias{TiDES_2025} selection criteria to the LSST DDF survey simulation. Note that the distributions are normalised and that the phase has been truncated at 50 days.}
    \label{fig:ddf_91bg_phase_distros}
\end{figure}

\begin{figure}
    \centering
    \includegraphics[width=\linewidth]{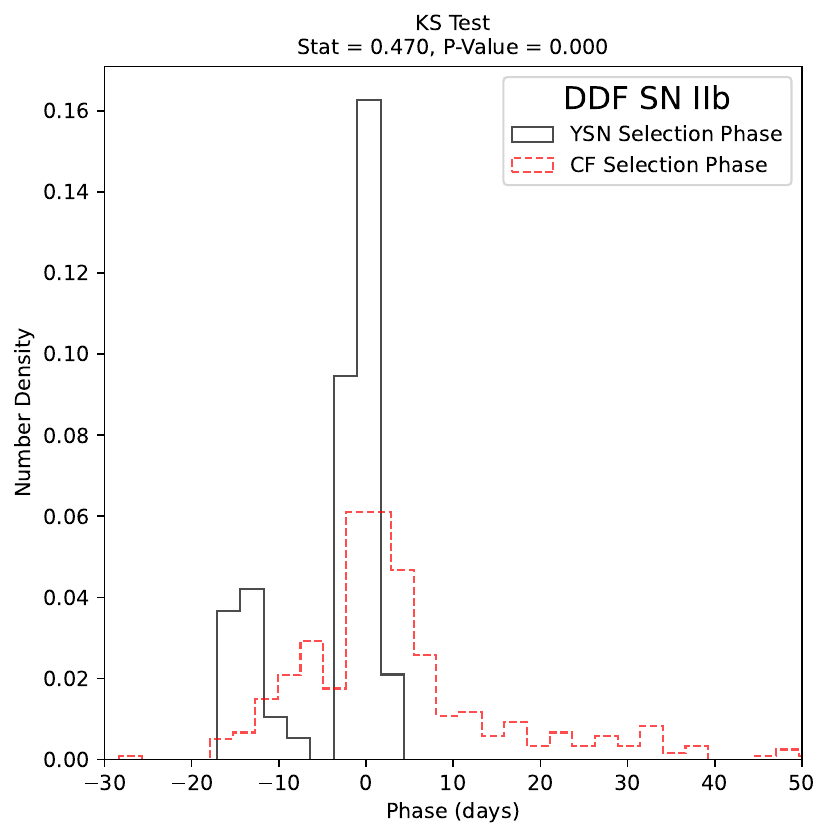}
    \caption{Comparison between the SN IIb selection phase distributions produced by applying our selection criteria (YSN; see Table \ref{tab:lsst_selection_criteria}) and the \citetalias{TiDES_2025} selection criteria to the LSST DDF survey simulation. Note that the distributions are normalised and that the phase has been truncated at 50 days.}
    \label{fig:ddf_iib_phase_distros}
\end{figure}

To investigate if our selection criteria selects YSNe, we present in Figures \ref{fig:ddf_ia_phase_distros}, \ref{fig:ddf_91bg_phase_distros}, and  \ref{fig:ddf_iib_phase_distros} the selection phase distributions for SNe Ia, 91bg, and IIb that were produced by our selection criteria and the \citetalias{TiDES_2025} criteria.
We present in Appendix \ref{app:ddf_phase_distros} the phase distributions for the other classes of transients. However, it should be noted that the classes presented in Appendix \ref{app:ddf_phase_distros} suffer from low number statistics and are only included for completeness.

Figures \ref{fig:ddf_ia_phase_distros} and \ref{fig:ddf_91bg_phase_distros} show that for SNe Ia and 91bg-like SNe (a subset of SNe Ia) the selection phase distributions produced by our selection criteria are unimodal with peaks that occur approximately 5 to 10 days before those produced by the \citetalias{TiDES_2025} selection criteria. Furthermore, our selection criteria is shown to only select pre-peak SNe Ia and 91bg-like SNe, with a maximum selection phase of approximately $-5$ days. 
The two-sided KS tests performed on the distributions for the SNe Ia and 91bg both returned P-values less than 0.0005, indicating that the selection phase distributions of the YSN candidate and \citetalias{TiDES_2025} samples are drawn from different distributions.  

In contrast, the SNe IIb selection phase distributions, presented in Figure \ref{fig:ddf_iib_phase_distros}, show that our selection criteria produces a selection phase distribution that is bimodal, with the dominant peak occurring at a phase of ${\sim}0$, which is approximately the same as the selection phase distribution produced by the \citetalias{TiDES_2025} selection criteria. The secondary peak of the distribution occurs at a phase of ${\sim}-15$ days. As with the WFD SNe IIb, the cause of this bimodality unclear.
The distribution displayed in Figure \ref{fig:ddf_iib_phase_distros} also shows that our selection criteria selects some (18 from Table \ref{tab:ddf_samples}) post-peak SNe IIb, with a maximum selection phase of ${\sim}4$ days.

As with the LSST WFD survey samples, to understand the crossover between the objects selected by our YSN selection criteria and the \citetalias{TiDES_2025} selection criteria, we present in Table \ref{tab:ddf_ysne_tides_objects_numbers} the number of selected transients that are common to both sets of selection criteria. We also include the number of transients that are only selected by our selection criteria.
As is shown, there are $548$ SNe Ia and $239$ non-SNe Ia that are commonly selected. Our YSN selection criteria selects only $8$ transients that were not selected by the \citetalias{TiDES_2025} selection criteria.

\begin{table}
        \caption{Number of LSST DDF simulated transients that were commonly selected by both our YSN selection criteria (see Table \ref{tab:lsst_selection_criteria}) and the \citetalias{TiDES_2025} criteria. Also included is the number of simulated transients that were only selected by our YSN criteria. Additionally, provided for the commonly selected transients are the average phases of selection by the \citetalias{TiDES_2025} criteria and our YSN criteria.}
    \label{tab:ddf_ysne_tides_objects_numbers}
        \centering
    \begin{tabular}{@{\hskip 3pt \vline \hskip 3pt} 
                c @{\hskip 3pt \vline \hskip 3pt} 
                c @{\hskip 3pt \vline \hskip 3pt}
                r |
                r @{\hskip 10pt \vline \hskip 3pt} 
                c @{\hskip 3pt \vline \hskip 3pt}}
    \hline
    \hline
    Classification 
        & \multicolumn{3}{c @{\hskip 3pt \vline \hskip 3pt}}{Commonly Selected} 
        & \multicolumn{1}{c @{\hskip 3pt \vline \hskip 3pt}}{YSN Uniquely} \\
    
        & \multicolumn{1}{c @{\hskip 3pt \vline \hskip 3pt}}{Quantity} 
        & \multicolumn{2}{c @{\hskip 3pt \vline \hskip 3pt}}{Mean Selection Phase} 
        & \multicolumn{1}{c @{\hskip 3pt \vline \hskip 3pt}}{Selected} \\

        &
        & \multicolumn{1}{c @{\hskip 10pt}}{\citetalias{TiDES_2025}} 
        & \multicolumn{1}{c @{\hskip 3pt \vline \hskip 3pt}}{YSN} 
        & \multicolumn{1}{c @{\hskip 3pt \vline \hskip 3pt}}{} \\
    \hline
    \hline
            Ia & 548 & $-13$ d & $-13$ d & 2  \\
            Iax & 6 & $-13$ d & $-14$ d & 0  \\
            91bg & 31 & $-8$ d & $-9$ d & 0  \\
            Ib & 5 & $-10$ d & $-10$ d & 0  \\
            Ic & 8 & $-10$ d & $-10$ d & 1  \\
            Ic-BL & 8 & $-7$ d & $-9$ d & 1  \\
            II & 57 & $-1$ d & $-2$ d & 1  \\
            IIb & 69 & $-3$ d & $-4$ d & 2  \\
            IIn & 47 & $-5$ d & $-5$ d & 1  \\
            SLSN & 4 & $-36$ d & $-36$ d & 0 \\
            CART & 1 & $-3$ d & $-3$ d & 0  \\
            TDE & 3 & $-27$ d & $-27$ d & 0  \\
        \hline
            Total & 787 & & & 8 \\
        \hline
        \hline 
    \end{tabular}
\end{table}

\begin{figure}
    \centering
    \includegraphics[width=\linewidth]{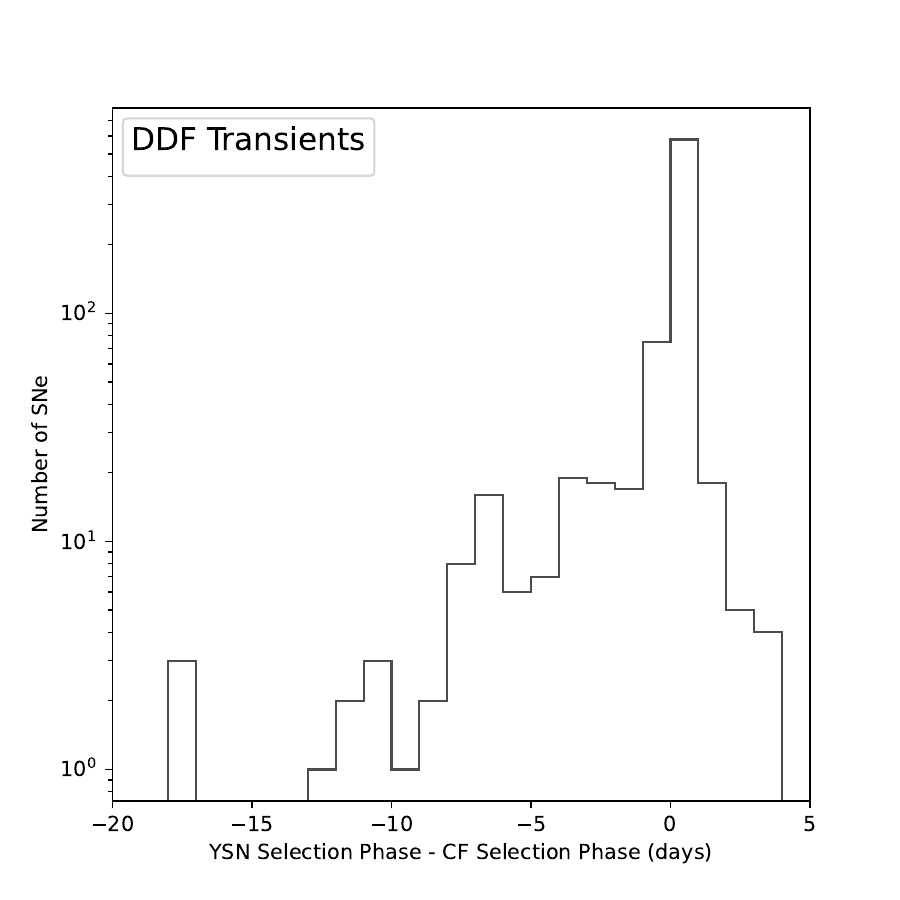}
    \caption{Distribution of the time differences of the selection of the LSST DDF transients that were selected by both our YSN selection criteria and the \citetalias{TiDES_2025} criteria. Note that all of the simulated transient types (see Table \ref{tab:wfd_samples}) are included.}
    \label{fig:ddf_common_trigger_diff}
\end{figure}

For the commonly selected LSST DDF survey transients, we directly compared the times at which they were selected by the two sets of selection criteria. Presented in Figure \ref{fig:ddf_common_trigger_diff} is the distribution of the time difference between when a commonly selected target was selected by our YSN selection criteria and the \citetalias{TiDES_2025} selection criteria. As indicated, approximately $600$ of the commonly selected transients have a positive time difference (selected by the \citetalias{TiDES_2025} criteria first) up to a maximum of 4 days. The remaining ${\sim}200$ commonly selected transients are shown to have a negative time difference (selected by YSN criteria first) as low as $-18$ days.

\subsection{YSN Selection Criteria Performance on LSST Simulations}\label{subsec:lsst_discussion}

\subsubsection{Sample Size}

From our results, presented in Section \ref{subsec:lsst_results}, it is evident from Tables \ref{tab:wfd_samples} and \ref{tab:ddf_samples} that our selection criteria produces a sample of transients that is not insignificant in number. From the 5 year LSST simulation, our YSN criteria selected a total of $56408$ WFD survey transients and $694$ DDF survey transients for follow-up observations. However, our YSN candidate samples are much smaller than the samples produced by the \citetalias{TiDES_2025} selection criteria, with their WFD and DDF samples containing ${\sim}14$ and ${\sim}11$ times more transients respectively.
Although our YSN candidate sample of selected transients is smaller than the \citetalias{TiDES_2025} sample, the aim of our selection criteria is not to replace the current selection criteria of TiDES (\citetalias{TiDES_2025} criteria) or of a TiDES-like survey. Instead our YSN selection criteria are to be used in conjunction with the pre-existing selection criteria, with the aim of selecting some of the transients at earlier phases to enhance the YSN samples of TiDES-like surveys. With this in mind, the size of our YSN candidate sample indicates that incorporating our selection criteria into a TiDES-like survey has the potential to have a non-insignificant affect on the resulting SN sample.

\subsubsection{Early Selection Effectiveness}

Our results presented in Section \ref{subsec:lsst_results}, specifically Figures \ref{fig:wfd_ia_phase_distros}-\ref{fig:wfd_ii_iib_iin_ic-bl_phase_distros} and \ref{fig:ddf_ia_phase_distros}-\ref{fig:ddf_iib_phase_distros}, showed that our selection criteria can select SN at earlier phases than the \citetalias{TiDES_2025} selection criteria. From Figure \ref{fig:wfd_ia_phase_distros}, it is shown that for the LSST WFD SNe Ia the selection phase distribution produced by our selection criteria peaks approximately 10 days before the selection phase distribution produced by the \citetalias{TiDES_2025} selection criteria. This general trend is seen for all of the transient classes across both the WFD and DDF surveys (where there is no low number statistics), which indicates that our selection criteria on average selects younger objects that the \citetalias{TiDES_2025} criteria, therefore enhancing the the early SN sample in the TiDES transient programme.

However, there are a two transient classes, SLSN and SN IIn, where the \citetalias{TiDES_2025} selection criteria can select the SN at earlier phases than our YSN criteria. While the phase distributions for these SN classes produced by our selection criteria have a peak at earlier phases, the \citetalias{TiDES_2025} criteria is able to select SLSN and SN IIn over 50 days earlier than our YSN criteria. This is demonstrated in Figures \ref{fig:wfd_slsn_phase_distros} and \ref{fig:wfd_ii_iib_iin_ic-bl_phase_distros} (bottom left). For example, for the WFD survey SLSNe, the \citetalias{TiDES_2025} selection criteria was able to select them as early as ${\sim}200$ days before peak while our YSN selection criteria only selected them as early as ${\sim}100$ days before peak.
However, these early detections are unrealistic, likely resulting from the over-extrapolation of the spectrophotometric templates used in the simulation of their LSST photometry.
Although an artefact of the simulations, the earlier selection of transients by the \citetalias{TiDES_2025} criteria highlights that our selection criteria may not be optimised for each transient class. Specifically, transient types that exhibit an initial relatively slow phase of brightening in their early evolution, such as SLSNe, will not be selected at these early phases by our YSN selection criteria due to the brightening rate criterion, which is designed to reject slowly brightening sources.

To gain a better understanding of how our YSN selection criteria performed, we compared the transients that were commonly selected by both the YSN and \citetalias{TiDES_2025} selection criteria.
As shown by Figures \ref{fig:wfd_common_trigger_diff} and \ref{fig:ddf_common_trigger_diff}, our YSN selection criteria was able to select ${\sim}35000$ WFD survey transients and ${\sim}200$ DDF survey transients at earlier times than the \citetalias{TiDES_2025} criteria. On average the commonly selected transients were selected by our YSN selection criteria three to five days earlier than the \citetalias{TiDES_2025} criteria, as is demonstrated by Table \ref{tab:wfd_ysne_tides_objects_numbers}. Some of these transients were selected over 100 days earlier by our YSN selection criteria. This demonstrates that our YSN selection criteria can prove very useful for selecting transients at phases earlier than is currently possible with the \citetalias{TiDES_2025} criteria.

On the contrary, the \citetalias{TiDES_2025} selection criteria selected ${\sim}19000$ WFD and ${\sim}600$ DDF survey transients earlier than our YSN selection criteria. However, upon further investigation, most of these transients (with the exception of ${\sim}4000$) were selected within the same night but from observations taken at different times. Considering that TiDES-like surveys will not be able to follow-up a selected transient within the same night, due to there being no target of opportunity mode, a few hours difference between the YSN and \citetalias{TiDES_2025} selection criteria triggers will have no effect on when the selected SN could potentially be followed-up by a TiDES-like survey.

Investigating the remaining ${\sim}4000$ transients, it was found that the \citetalias{TiDES_2025} criteria selected many of them before our YSN criteria because they lacked multiple observations in same filter. At least two observations in a filter are required to calculate the brightening rate used in the YSN selection criteria, whereas the \citetalias{TiDES_2025} criteria does not require multiple observations in one filter and so can select some targets earlier than our YSN criteria.  
To solve this issue the brightening rate criterion could be removed from our selection criteria. However, the brightening rate criterion is used to limit the contamination that is selected. To retain the brightening rate and to allow for selection based on two detections in any filter, future works could investigate the use of cross-filter brightening rates.

Considering the improvements in earlier selection of ${\sim}20000$ SNe, we believe that our YSN selection criteria is suitable to be implemented in to a TiDES-like survey. However, we must finally consider the potential contamination of the YSN candidate sample in order to determine if we should implement our YSN selection criteria into TiDES or a TiDES-like survey.

\subsubsection{Sample Contamination}\label{subsec:lsst_discussion_contamination}

As previously discussed in Section \ref{subsubsec:ztf_evalutation}, our YSN selection criteria should have a sufficiently high purity (low contamination). From our ZTF YSN candidate sample, we calculated that the non SN contamination of the sample was $23$ percent. However, we could not conclude if this level of contamination was sufficiently low to not waste too many fibre hours. By extrapolating the level of contamination from the ZTF YSN candidate sample to the LSST YSN candidate sample, we can estimate how many fibre hours a TiDES-like survey will potentially waste on non-SN sources.

From Tables \ref{tab:wfd_samples} and \ref{tab:ddf_samples}, it is shown that the selected YSN candidate sample (WFD and DDF samples combined) contains $56951$ SNe (exclusion of TDE).
Assuming that the contamination of the ZTF and LSST samples will be comparable, the number of non SN contaminants that would be selected as part of the LSST YSN candidate sample is $17011$ objects ($56951 \times 0.23 / (1-0.23)$).

To estimate how many of the contaminants could be observed by a 4MOST/TiDES-like survey, we begin by making a few assumptions: 4MOST observes for on average 9 hours per night \citep{Guiglion_2019}, an average 4MOST pointing lasts 1 hour, and 15 SNe are observed per 4MOST pointing. Based on these assumptions, a 4MOST-like survey can observe an estimated 135 SNe per night. 4MOST is a 5 year long survey, however, approximately 300 nights per year at Paranal are usable for observations \citep{Guiglion_2019}, resulting in an estimated 1500 observable nights over the lifetime of 4MOST. As a result, an optimistic estimate suggests that a 4MOST-like survey could observe up to $202,500$ SNe over its lifetime.
This is much less than the total number of unique transients that are selected by our criteria and the \citetalias{TiDES_2025} criteria for follow-up with a 4MOST-like survey, which is approximately $780,000$ transients. Therefore, up to ${\sim}26$ percent of the selected samples can be estimated to be observed by a 4MOST-like survey, which means that ${\sim}4423$ ($17011 \times 0.26$) contaminants selected by our YSN selection criteria are observed over the lifetime of a TiDES-like survey.

By assuming that each observed object receives 40 minutes of observation time, over the lifetime of a TiDES-like survey (5 years) the fibre hours wasted on the non-SN objects is ${\sim}2949$ hours ($40\,\rm{mins} \times 4423$). This equates to approximately $1.2$ percent of TiDES's total available fibre hours (250000 hours \citepalias{TiDES_2025}) or $2.9$ percent of the fibre hours allocated for the TiDES SN survey (100000 fibre hours, $40$ percent of the entirety of TiDES).
We note here that these estimates are lower limits. This is due to our assumption that the contamination rate between LSST and ZTF will be comparable. Although this might be true in the later years of LSST, during the early stages the contamination rate will be elevated due to contaminating transients, such as AGN and CVs, being discovered for the first time.
Based on our estimates of wasted fibre hours, we consider our selection criteria capable of producing a pure enough sample as to not waste too many fibre hours on non-SN targets.

\section{Optimal 4MOST-like Strategy}\label{sec:optimal_strategy}

The TiDES survey will spectroscopically follow-up transients that are selected from the LSST live transient alerts using 4MOST. As of now, there is no coordination between the observing strategies of LSST and 4MOST, which results in an unknown variable delay between the selection of an LSST SN and its follow-up 4MOST observation. This leads to many cases where the signal to noise ratio (SNR) of a 4MOST observed SN spectrum is much lower than its highest obtainable SNR. 
Furthermore, the lack of coordination also negatively impacts our YSN selection criteria, as the unknown variable delay could be days, weeks, or even months, at which point a selected YSN has evolved and is likely no longer an early-time SN.  
To fully utilise our YSN selection criteria, follow-up of the targets should be conducted immediately. However, this is not possible with a 4MOST-like instrument due to the lack of a target of opportunity mode. Immediate follow-up of some targets could be conducted with dedicated target of opportunity programmes on instruments such as X-shooter \citep{Vernet_2011} or {\sl Son-of-X-shooter} \citep[SOXS;][]{Schipani_2016}. However, these programmes would not provide the scale of observations that a TiDES-like survey could provide. Therefore, we propose to construct a 4MOST-like observing strategy for the case of maximising transient follow-up.

To optimise a 4MOST-like observing strategy for the case of a TiDES-like survey and our YSN selection criteria, we propose that the 4MOST-like strategy follows (with some delay) the LSST observing strategy. In theory, this strategy would result in our LSST selected targets always being spectroscopically observed soon after their selection. This should increase the number of observed targets, improve the SN spectra SNR, and aid in obtaining early-time SN spectra. 
There are of course caveats to consider regarding the feasibility of a 4MOST-like survey following the strategy of LSST. For example, LSST will observe the southern sky approximately every 3 days\footnote{\url{https://survey-strategy.lsst.io/baseline/wfd.html}}, which is not possible with a survey such as 4MOST, resulting in some LSST fields having to be skipped.
As such, in this work we do not seek to construct a full 4MOST-like observing strategy, but instead we investigate and suggest guidelines for future works towards designing and simulating a full 4MOST-like observing strategy for time-domain science.
Furthermore, we only investigate 4MOST-like strategies that cover the LSST WFD fields, as in the case of 4MOST the DDF fields will have a cadence based observing strategy based on requirements from the TiDES reverberation mapping survey \citepalias{TiDES_2025} and other 4MOST surveys.

\subsection{Investigating Observing Strategies}\label{subsec:obs_strat_investigation_methods}

\subsubsection{Simulating 4MOST-Like SN Spectra}\label{subsubsec:sim_4most_spec}

To simulate a SN spectrum as observed by a 4MOST-like instrument, for a given SN in the LSST simulation that was used in Section \ref{sec:lsst_application}, we first created its template spectrum at a given phase. This was achieved by using the \texttt{Python} package \texttt{SNCosmo} \citep{sncosmo}, which can extract a template spectrum from a spectrophotometric model for the given input parameters. Different models have different input parameters, for example SN Ia simulated with the extended SALT2 model \citep{Hounsell_2018} require the parameters of redshift, phase, SALT2 colour parameter, SALT2 stretch parameter, and apparent magnitude. Whereas a SN II simulated using the models of \citet{vincenzi_2019} requires only phase, redshift, and apparent magnitude. The spectrophotometric model and its input parameters for each of the SN in the LSST simulation are recorded in the simulation output files.

To the template spectra, we also applied the effect of Galactic extinction, which was achieved by using the \texttt{SNCosmo} method \texttt{``F99Dust''} that applies the extinction model of \citet{Fitzpatrick_1999}. We used the total-to-selective extinction ratio $\rm{R}_{\rm{V}}=3.1$ and the extinction values defined for each SN in the LSST simulation output files.
Although we included Galactic contamination in our SN spectral templates, we did not attempt to include contamination from the SNe's host galaxy. 

With the template spectra created, we then simulated the effects of observing the SN spectra with 4MOST, which was accomplished by using the 4MOST exposure time calculator (ETC)\footnote{We used the \texttt{Python} API of the 4MOST ETC (\url{https://escience.aip.de/readthedocs/OpSys/etc/master/index.html}).}.
As we simulated observed spectra for a TiDES-like survey, we used the low-resolution spectrograph as this will be used by TiDES during operations \citepalias{TiDES_2025}. For the observing conditions, we assumed the following typical observing conditions for TiDES targets: zenith angle of $45\deg$, seeing of $0.8$ arcsec, grey sky brightness, and an exposure time of 40 minutes.
Although having constant observing conditions for all targets is unrealistic, varying the observing conditions would require development of a full 4MOST-like observing strategy, which is outside the scope of this work.

\subsubsection{Applying an Observing Strategy}

As previously mentioned, we investigated 4MOST-like WFD observing strategies that follow LSST with a delay time.
To accomplish this, we applied our methods presented in Section \ref{subsubsec:sim_4most_spec} to the WFD \citetalias{TiDES_2025} and WFD YSN candidate samples produced and presented in Section \ref{sec:lsst_application}, simulating their 4MOST observed spectra.
To apply the 4MOST-like observing strategies with different delay times, we altered the phases at which the template spectra were created for.
The phase that was used to produce a given SN's template spectrum is given as the phase at which the SN was selected plus the time delay of the 4MOST-like observing strategy. For example, an LSST SN selected at a phase of $-10$ days would be observed by a 4MOST-like strategy with a 3 day time delay at a phase of $-7$ days.
Using these methods, we simulated the spectra for the LSST WFD selected \citetalias{TiDES_2025} and YSN candidate samples as if they had been observed using 4MOST-like observing strategies with delay times of 1, 3, 5, and 7 days.

\subsubsection{Investigating Observing Strategy Effects}

To investigate the effects that the different time delays of the 4MOST-like observing strategy have on the resulting spectroscopically observed SN sample, we inspected the SNR of the spectra.
Specifically, we analysed the observed SN samples' SNR distributions as well as the percentage of simulated SNe whose spectra exceed certain SNR thresholds. Following \citetalias{TiDES_2025} we chose to define the SNR of a spectrum as the mean SNR in $15$\AA\ bins over the observer frame wavelengths $4500 - 8000$\AA, hereafter denoted as \snr. \snr thresholds of 5 and 3 were chosen as they can be used as a proxy for how reliably a SN spectrum can be classified. \citet{Balland_2009} showed that spectra with a \snr $>5$ provides reliable classifications, while a possible classification was achievable for spectra with a \snr as low as 3.
For the purpose of our study, using the SNR as a proxy for the reliability of the SN classifications is satisfactory. However, for a more in depth analysis of the SN classification reliability of 4MOST-like spectra, we refer the reader to \citet{Milligan}. 
Although the chosen SNR thresholds are a good proxy for spectra classifications, we also kept in mind that we want to maximise the spectra's SNR in order to reliably extract the spectral information required for performing astrophysical studies of SN.

\subsection{Observed SN Samples}\label{subsec:obs_strat_results}

\subsubsection{\citetalias{TiDES_2025} Sample}\label{subsecsec:obs_strat_results_tides}

\begin{figure}
    \centering
    \includegraphics[width=\linewidth]{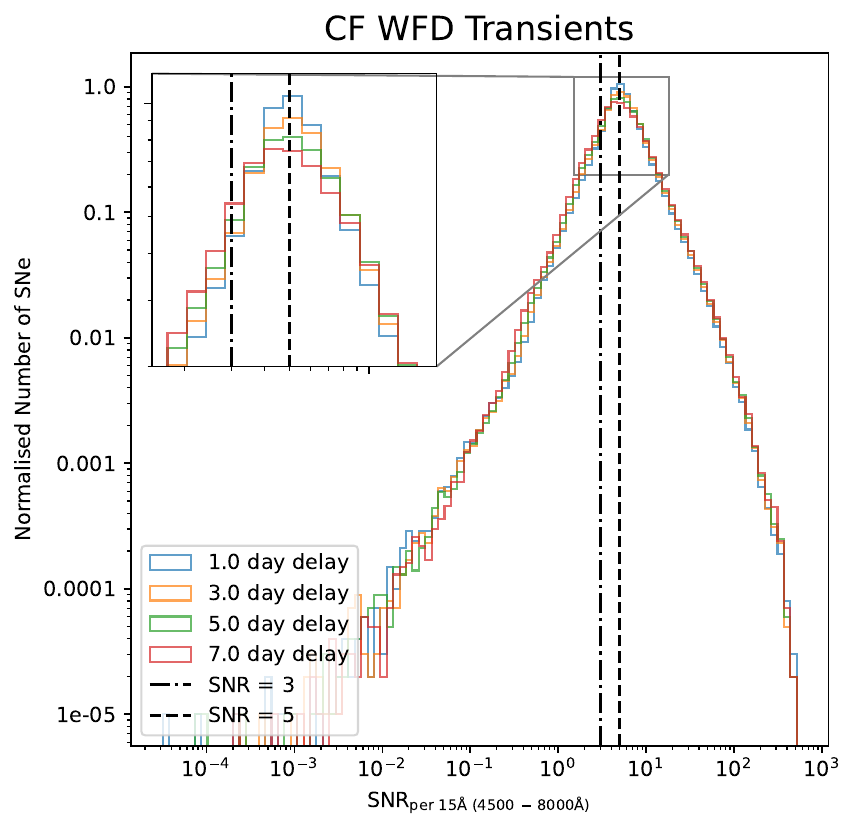}
    \caption{The \citetalias{TiDES_2025} LSST selected WFD SN sample's observed spectral \snr distributions. The SNe were observed using 4MOST observing strategies that followed the LSST strategy by 1, 3, 5, and 7 days.}
    \label{fig:tides_obs_snr_distros}
\end{figure}

Following the methods of Section \ref{subsec:obs_strat_investigation_methods}, we present in Figure \ref{fig:tides_obs_snr_distros} the SNR distributions of the observed \citetalias{TiDES_2025} WFD SN sample's spectra following observations with the 4MOST-like strategies that follow the strategy of LSST with a 1, 3, 5, or 7 day delay.
As is shown, the different observing strategies all produce similar SNR distributions that have no significantly notable differences. The SNR distributions of the 4MOST-like observed spectra all peak at a \snr of 5 regardless of the 4MOST-like observing strategy.

In addition to the \snr distributions, we present in Table \ref{tab:num_4most_observed} the percentage of transients whose observed 4MOST-like spectra exceed a \snr of 5 or 3 when observed using the 4MOST-like observing strategies that follow the LSST strategy with a 1, 3, 5, and 7 day delay. 
From the table it is shown that of the SNe observed under any of the investigated 4MOST-like observing strategies, more than $47.7$ percent ($75.2$ percent) of the observed \citetalias{TiDES_2025} WFD SNe have a \snr $>5$ ($>3$).
Table \ref{tab:num_4most_observed} also demonstrates that as the time delay between the LSST and 4MOST-like observing strategies increases, the percentage of observed \citetalias{TiDES_2025} WFD SNe with a \snr $>3$ decreases. This trend is not seen for the percentage of observed \citetalias{TiDES_2025} WFD SNe with a \snr $>5$, as the maximum percentage occurs with the 3 day delayed 4MOST observing strategy. The percentage of observed \citetalias{TiDES_2025} WFD SNe with a \snr $>5$ under the 3 day delayed strategy is $51.5$ percent, while the strategies with a 1, 3, and 7 day delays all have percentages that are more than $0.7$ percent less than the 3 day delay strategy.

\begin{table}
       \caption{For both the LSST selected WFD SN samples (\citetalias{TiDES_2025} WFD and YSN WFD), presented are the percentages of 4MOST-like observed SNe whose spectra exceeded a \snr of 5 or a \snr of 3. These results are provided for the 4MOST-like observing strategies that follow the LSST strategy by delays of 1, 3, 5, and 7 days.} 
    \label{tab:num_4most_observed}
    \centering
    \setlength{\tabcolsep}{2.5pt}
    \begin{tabular}{@{\hskip 5pt \vline \hskip 5pt} c @{\hskip 5pt \vline \hskip 5pt} c | c @{\hskip 5pt \vline \hskip 5pt} c | c @{\hskip 5pt \vline \hskip 5pt} c | c @{\hskip 5pt \vline \hskip 5pt}}
        \hline
        \hline
             & \multicolumn{4}{c @{\hskip 5pt \vline \hskip 5pt}}{WFD SN Sample} \\
            
            4MOST-Like & \multicolumn{2}{c @{\hskip 5pt \vline \hskip 5pt}}{\citetalias{TiDES_2025}} & \multicolumn{2}{c @{\hskip 5pt \vline \hskip 5pt}}{YSN Candidate} \\
            
            Observing Strategies & \multicolumn{1}{c}{SNR $>3$} & \multicolumn{1}{c @{\hskip 5pt \vline \hskip 5pt}}{SNR $>5$} & \multicolumn{1}{c}{SNR $>3$} & \multicolumn{1}{c @{\hskip 5pt \vline \hskip 5pt}}{SNR $>5$} \\
        \hline
        \hline

            1 Day Delay & $81.2\%$ & $50.8\%$ & $94.1\%$ & $75.0\%$ \\
            3 Day Delay & $79.7\%$ & $51.5\%$ & $95.1\%$ & $85.6\%$ \\
            5 Day Delay & $77.8\%$ & $50.1\%$ & $94.9\%$ & $87.5\%$ \\
            7 Day Delay & $75.2\%$ & $47.7\%$ & $94.2\%$ & $87.7\%$ \\

        \hline
        \hline 
    \end{tabular}
\end{table}

\subsubsection{YSN candidate Sample}\label{subsecsec:obs_strat_results_ysn}

\begin{figure}
    \centering
    \includegraphics[width=\linewidth]{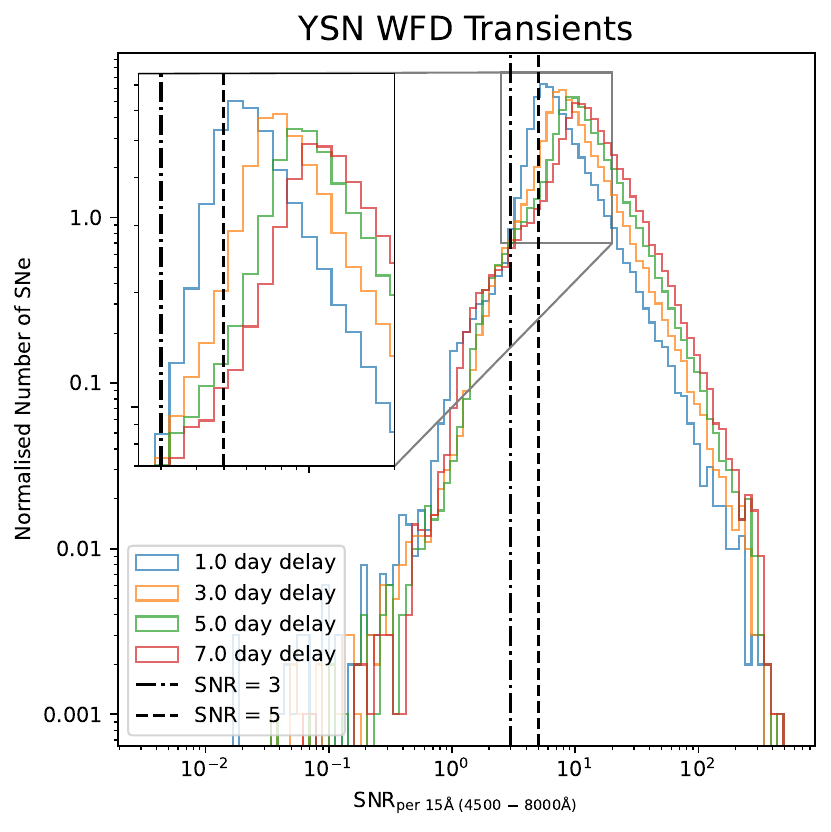}

    \caption{The LSST selected YSN candidate sample's observed spectral \snr distributions. The SNe were observed using 4MOST observing strategies that followed the LSST strategy by 1, 3, 5, and 7 days.}
    \label{fig:ysn_obs_snr_distros}
\end{figure}

As with the \citetalias{TiDES_2025} sample, we present in Figure \ref{fig:ysn_obs_snr_distros} the SNR distributions of the observed WFD YSN candidate sample's spectra following observations with the 4MOST-like strategies that follow the strategy of LSST with a 1, 3, 5, or 7 day delay.
As is shown, the different observing strategies all produce similarly shaped SNR distributions. The peaks of the SNR distributions for the different observing strategies are shifted to higher SNR the larger the strategy's time delay is.
For example, the WFD YSN candidate sample observed using the 7 day delayed 4MOST-like strategy peaks at a \snr of ${\sim}10$, while the 1 day delayed strategy peaks at a \snr of ${\sim}6$.

In addition, we also present in Table \ref{tab:num_4most_observed}, the percentage of transients whose observed 4MOST-like spectra exceed a \snr of 5 or 3 when observed using the 4MOST-like observing strategies that follow the LSST strategy with a 1, 3, 5, and 7 day delay. 
As the Table shows, the percentage of observed YSN WFD SN spectra whose \snr is greater than $5$ increases as the delay between the 4MOST-like and LSST observing strategies is increased, with $88.9$ percent of WFD YSN observed having a \snr $>5$ under the 7 day delayed 4MOST-like strategy.
In contrast, the percentage of observed WFD YSN whose spectra have a \snr $>3$ is at maximum when the 4MOST-like 3 day delayed strategy is applied.

\subsection{Discussion}

\subsubsection{\citetalias{TiDES_2025} Sample}

The results presented in Section \ref{subsecsec:obs_strat_results_tides}, showed that the percentage of observed \citetalias{TiDES_2025} WFD SN whose spectra have a \snr $>3$ decreases as the observing strategy time delay is increased from 1 day to 7 days. This suggests that a time delay of 1 day between the LSST and 4MOST-like observing strategies is the most optimal delay for obtaining the most spectra with a \snr $>3$.
On the contrary, the highest percentage of observed SN with a spectra whose \snr is $>5$ occurred when applying the 3 day delayed strategy, suggesting that this could be the more optimal strategy.
However, we must also consider the significance of the \snr thresholds of 5 and 3. As shown by \citet{Balland_2009}, SN spectra with a \snr $>5$ can be reliable classified, while classifications from spectra with a \snr as low as $3$ are possible but less reliable.

Taking the significance of the SNR thresholds into consideration, using a 1 day delayed 4MOST-like observing strategy provides the highest percentage of observed SN that are likely to be classified, 81.2 percent of the \citetalias{TiDES_2025} WFD sample. This is 1.5 percent more than observed with the 3 day delayed strategy.
However, the 3 day delayed 4MOST-like strategy provides the most spectra that could be reliably classified, 51.5 percent of the \citetalias{TiDES_2025} WFD SN sample, which is 0.7 percent more than observed with the 1 day delayed strategy.

For the case of the \citetalias{TiDES_2025} SN sample, to determine which strategy is optimal depends on if one values the prospect of obtaining more SN classifications over the reliability of the classifications. The TiDES SN survey values the reliability of the classifications and the physics that can be constrained from the spectra. Therefore, a 3 day delayed 4MOST-like observing strategy, which provides the highest percentage of spectra that can reliably classified (\snr $>5$), is more optimal for the \citetalias{TiDES_2025} SN sample.

\subsubsection{YSN Candidate Sample}

The results presented in Section \ref{subsecsec:obs_strat_results_ysn} showed that the SNR distribution is shifted to higher SNR as the delay time is increased. This can be explained by the fact that our YSN objects are predominantly selected at pre-peak phases. As the 4MOST-like observing strategy time delay is increased, the number of SNe that evolve to near peak brightness increases, resulting in an increase in the number of SNe that are bright enough such that their spectral \snr $>5$.
From the observing strategies investigated, the longest delay of 7 days produces the highest percentage (87.7 percent) of SNe that have good quality spectra (\snr $>5$). Although this strategy is the most optimal for obtaining good quality spectra, it is counterproductive for observing early-time SN spectra.

For a TiDES-like survey to fully capitalise on the earlier SN selection provided by our YSN selection criteria, quick follow up of the YSN targets is required. Ideally an event would be followed up within minutes of selection, however, this is not possible for a TiDES-like survey as there is no target of opportunity mode. Therefore, a 4MOST-like observing strategy that follows the LSST observing strategy with a delay of 1 day is most optimal for making the most out of our YSN selection criteria's early triggers.
One limitation of this strategy, as shown by our results in Section \ref{subsecsec:obs_strat_results_ysn}, is that it produces the lowest percentage of SNe with good quality spectra, at 12.7 percent less than the most optimal strategy (7 day delay strategy). However, the percentage of SNe for which a possible classification is achievable (spectrum SNR $>3$) is 94.1 percent, which is at most 1.0 percent less than the other observing strategies investigated, and only 0.1 percent less than the 7 day delayed observing strategy.

\section{Conclusions}\label{sec:conclusion}

We have developed a set of selection criteria for TiDES-like surveys that will select YSNe from the LSST live transient alerts for spectroscopic follow-up. The aim of our selection criteria was to enhance the YSN samples of TiDES-like surveys by selecting transient events sooner than a TiDES-like selection criteria, potentially allowing for observations of SN spectra at earlier phases.

To develop our selection criteria, we first utilised the transient alerts from ZTF, allowing us to develop a set of selection criteria that could produce a candidate sample of YSN (SNe selected before a phase of $-10$ days) that was not overly contaminated with non-SN transients, and spurious transient alerts. By applying our selection criteria to the ZTF alerts over a period of 60 nights, we produced a YSN candidate sample consisting of 60 classified YSNe and 17 unclassified but likely YSNe. The non SN contamination of our produced ZTF YSN candidate sample was $23$ percent.

To evaluate the effect that our selection criteria could have on a TiDES-like SN survey, we exploited the LSST simulations of \citetalias{TiDES_2025}, producing an LSST YSN candidate sample by applying the following ZTF developed YSN selection criteria: 

\begin{enumerate}
    \item Consider only LSST $griz$-bands.
    \item Object is brighter than 22.5 mag in any $griz$-bands.
    \item Object's declination is between $5$\degree and $-70$\degree. 
    \item Object's Galactic latitude is not between $-10$\degree and 10\degree.
    \item Object has two or more $>5\sigma$ detections in a given filter.
    \item Object has no previous detections more than 7 days before the latest detection.
    \item Brightening rate in any $griz$-band $> 0.2$ mag/day.    
\end{enumerate}

In total our YSN selection criteria produced an LSST selected SN sample consisting of 56408 WFD survey transients and 694 DDF survey transients. Although this sample is significantly smaller than the \citetalias{TiDES_2025} sample (857458 WFD survey and 9190 DDF survey transients), our selection criteria were developed to select early transients rather than produce a large sample, and is intended to be applied in conjunction with existing selection criteria such as those used by TiDES \citepalias{TiDES_2025}.
We demonstrated that our YSN selection criteria can provide earlier selection of LSST observed SNe, allowing TiDES-like surveys to enhance their early SN science capabilities.  

In addition, we also showed that our selection criteria is capable of producing a YSN candidate sample that is sufficiently pure. By extrapolating the contamination rate from the ZTF sample (23 percent) and estimating the maximum possible number of SNe that a TiDES-like survey could observe over its lifetime, we estimated (as a lower limit) that $2949$ fibre hours ($1.2$ percent of TiDES's total fibre hours) will be wasted over a 5 year survey. We believe that the number of fibre hours spent on non-SN sources is low enough such that it will not have a significant negative impact on a TiDES-like survey, especially when considering the earlier selection benefit that is gained by our YSN selection criteria.

Finally, we investigated different 4MOST-like observing strategies to optimise the output of a TiDES-like survey and our YSN criteria. Specifically, we investigated simplistic 4MOST-like observing strategies that follow the strategy of LSST with delays of 1, 3, 5, and 7 days, looking only at the LSST WFD fields.
For the \citetalias{TiDES_2025} sample, our results showed that the 3 day delayed 4MOST-like strategy was the most optimal strategy, as it provided the highest number of spectra (51.5 percent of observed WFD SNe) that can be reliably classified (\snr $>5$).
However, this was not replicated for the YSN candidate sample, as the 7 day delayed strategy provided the most spectra that can be reliably classified. We did not consider the 7 day delayed 4MOST-like strategy to be optimal for the YSN candidate sample due to the relatively long delay, which would counteract the early SN triggers. Instead, we considered the 1 day delayed strategy to be the most optimal strategy for the YSN candidate sample, as it provides the quickest follow-up. Although we showed that this strategy produced the lowest percentage of good quality spectra (spectra with a \snr $>5$), the percentage of good quality spectra is still relatively high at 75.0 percent. Therefore, the 4MOST-like observing strategy that follows the LSST strategy with a 1 day delay is most optimal for the YSN candidate sample and the resulting early-time SN science.   

In summary, this work has demonstrated the benefits that a TiDES-like survey can gain by implementing our YSN selection criteria along side its own selection criteria. Specifically, we have shown that our YSN criteria will enhance the early-time science capabilities of a TiDES-like survey.
Furthermore, we recommend that future work towards designing and simulating a 4MOST-like observing strategy should adopt a strategy that closely follows the LSST strategy with a delay of 3 days or a delay of 1 day for optimising the \citetalias{TiDES_2025} sample or the YSN candidate sample respectively.

\section*{Acknowledgements}

The authors thank Mathew Smith for their assistance with the simulations, Mark Sullivan for discussions and comments on earlier drafts of this work, and Alexander Fritz for their comments.
Extensive use of Python was used during this work, specifically the packages: NumPy \citep{harris_2020}, Astropy \citep{Astropy_2013, Astropy_2018, Astropy_2022}, and Matplotlib \citep{Hunter_2007}.
This work was supported by the Science and Technology Facilities Council (STFC) grant ST/X508810/1.
KM acknowledges funding from EU H2020 ERC grant no. 758638 and Horizon Europe ERC grant no. 101125877.
IH acknowledges support from STFC through grants ST/Y001230/1 and ST/V000713/1, and a Leverhulme Trust International Fellowship, reference IF-2023-027.
SJS acknowledges funding from STFC grants ST/Y001605/1, ST/X006506/1, ST/T000198/1, a Royal Society Research Professorship and the Hintze Charitable Foundation.

\section*{Data Availability}

All data are available upon reasonable request to the corresponding author.

\bibliographystyle{mnras}
\bibliography{ref}

\appendix

\section{LSST WFD Selection Phase Distributions}\label{app:wfd_phase_distros}

Presented are the WFD survey transients' selection phase distributions produced by applying our YSN selection criteria and the \citetalias{TiDES_2025} selection criteria to the simulated LSST WFD survey. Only the transient classes not presented in the main body are included here. Note that the distributions are normalised and that the phases have been truncated at 50 days.

\includegraphics[width=0.8\linewidth]{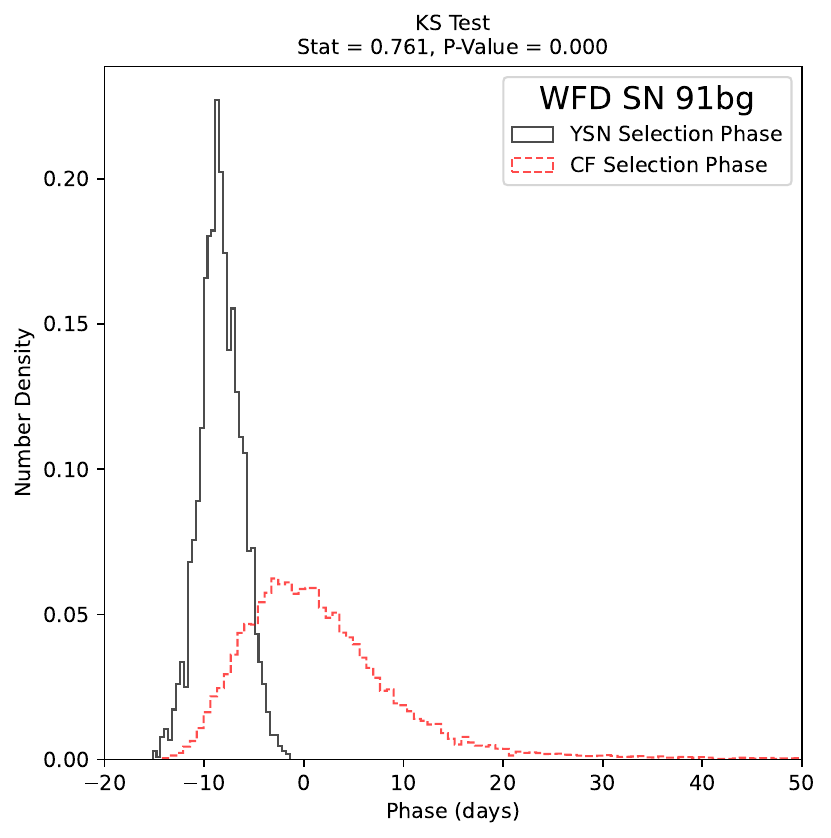}

\includegraphics[width=0.8\linewidth]{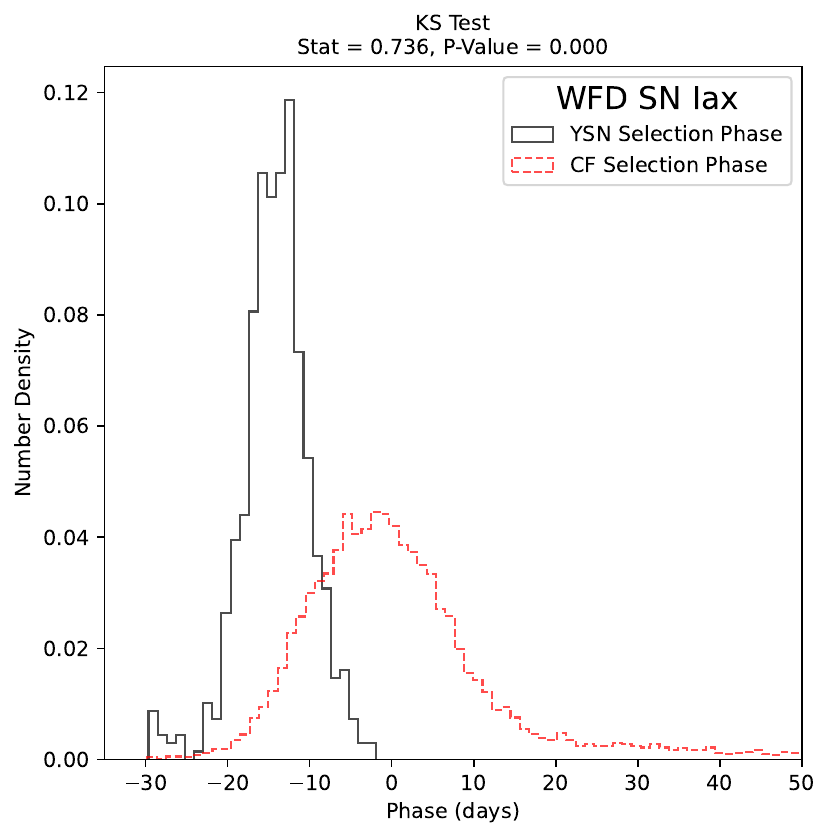}

\includegraphics[width=0.8\linewidth]{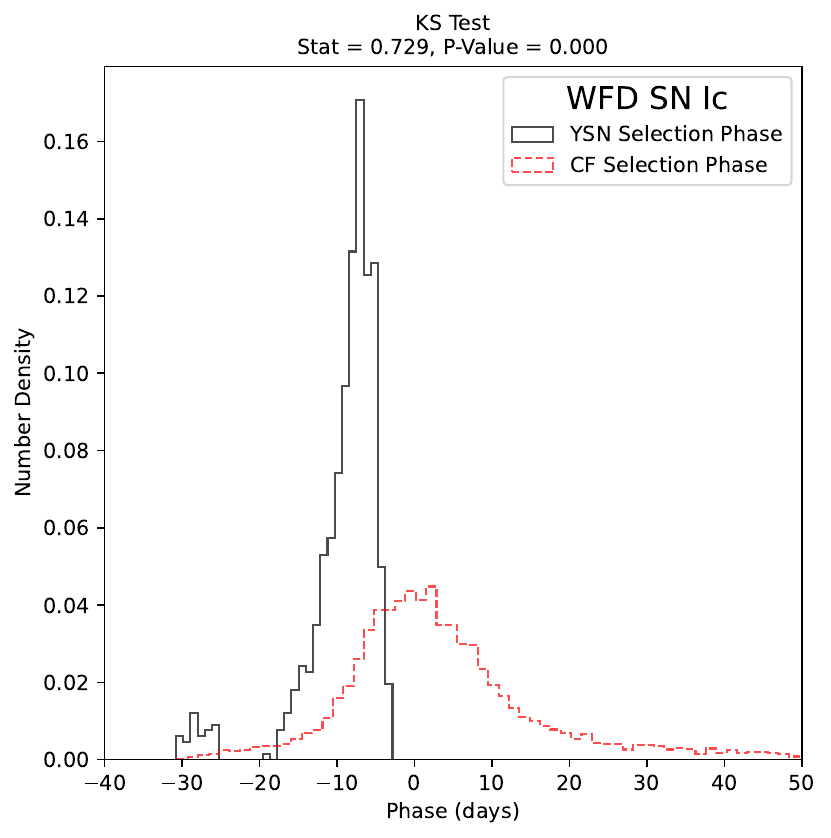} 

\includegraphics[width=0.8\linewidth]{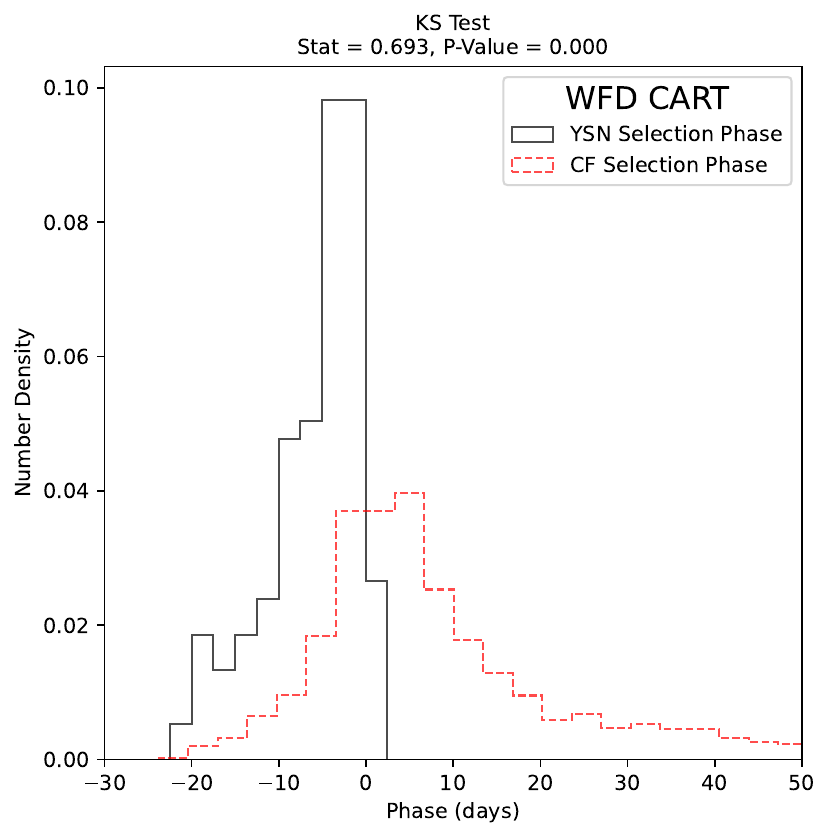}

\includegraphics[width=0.8\linewidth]{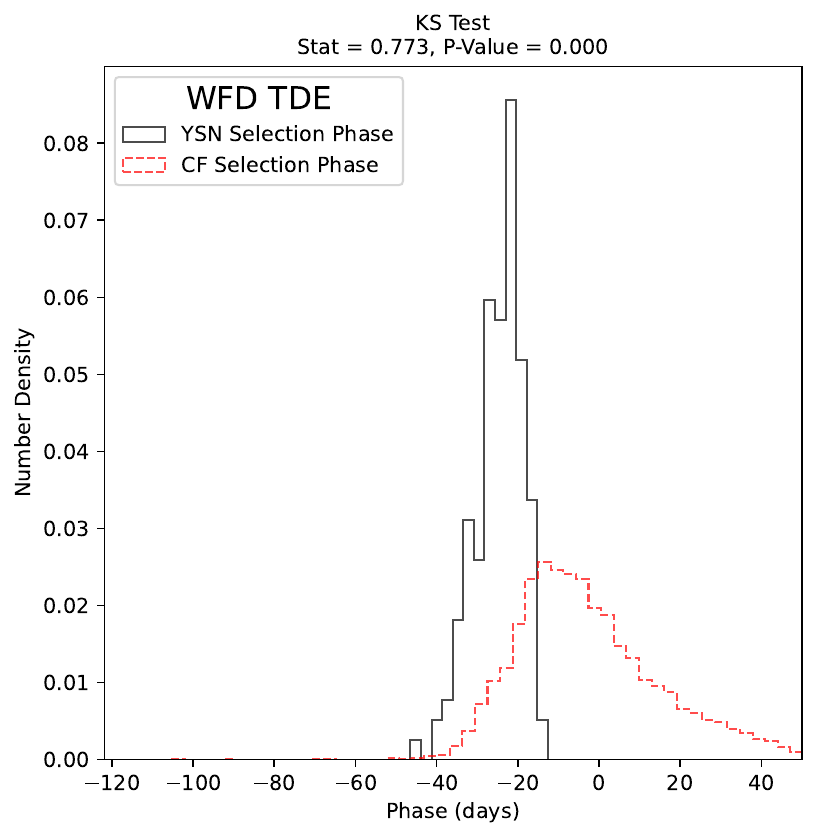}

\section{LSST DDF Selection Phase Distributions}\label{app:ddf_phase_distros}

Presented are the DDF survey transients' selection phase distributions produced by applying our YSN selection criteria and the \citetalias{TiDES_2025} selection criteria to the simulated LSST DDF survey. Only the transient classes not presented in the main body are included here. Note that the distributions are normalised and that the phases have been truncated at 50 days.

\includegraphics[width=0.8\linewidth]{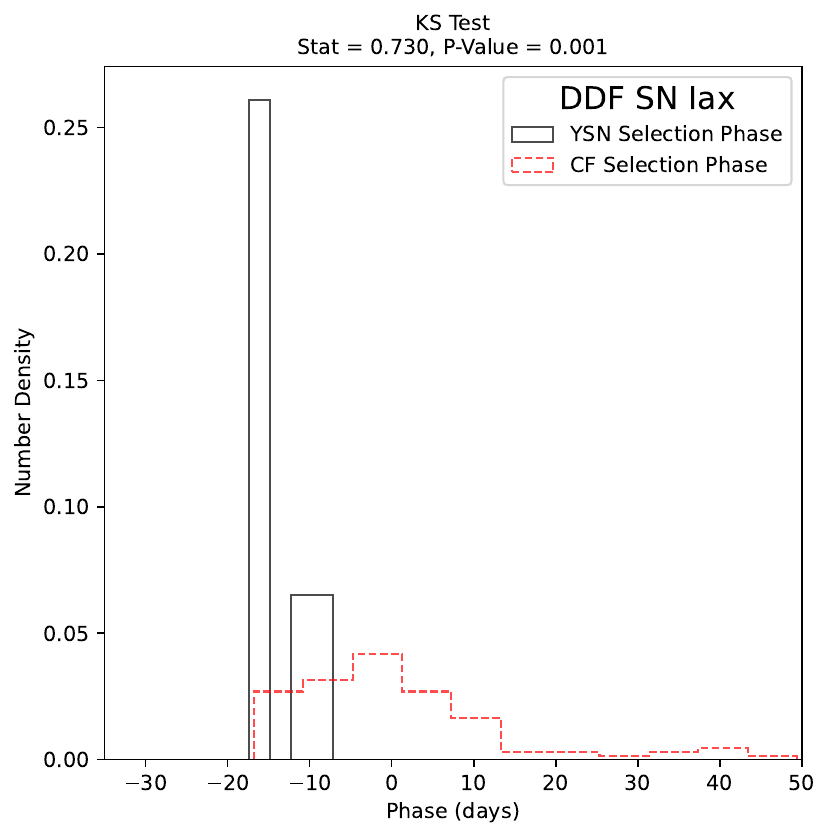}
\includegraphics[width=0.8\linewidth]{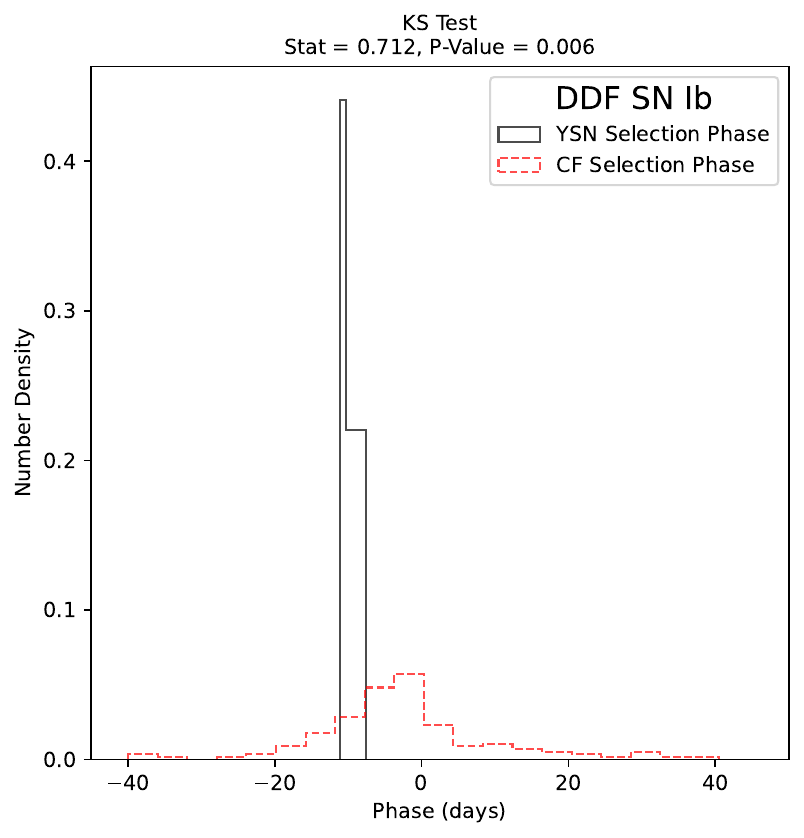}
\includegraphics[width=0.8\linewidth]{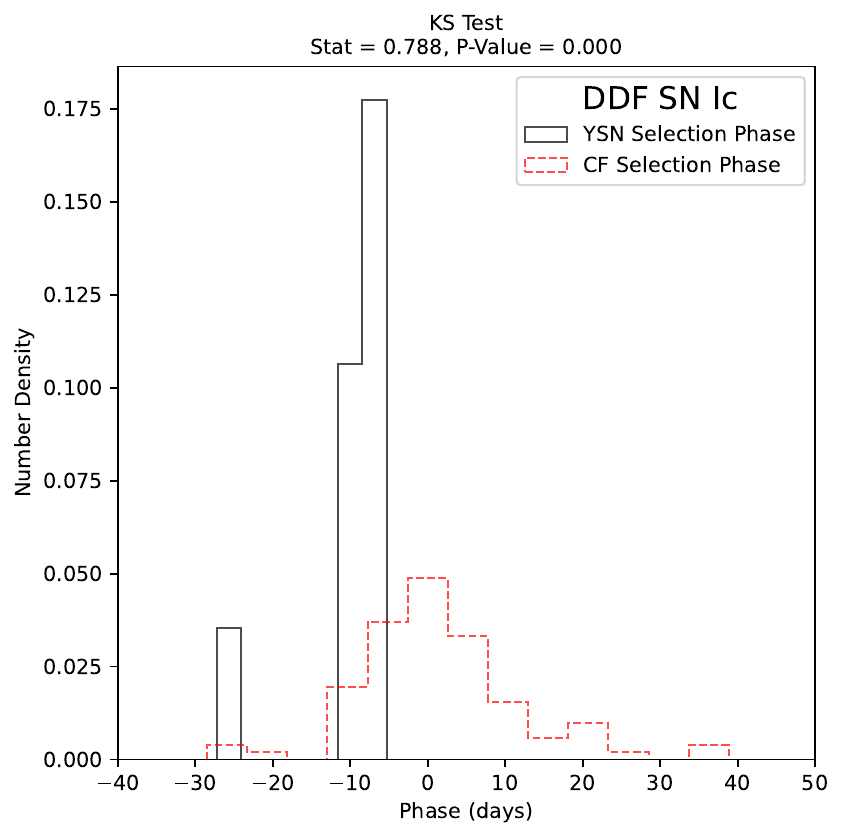}
\includegraphics[width=0.8\linewidth]{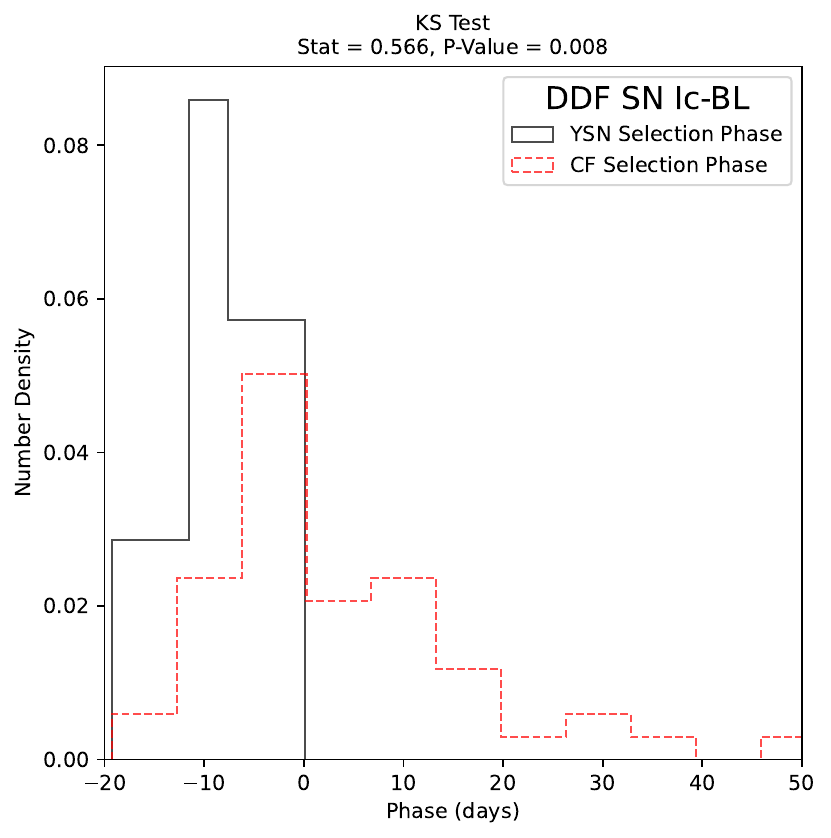}
\includegraphics[width=0.8\linewidth]{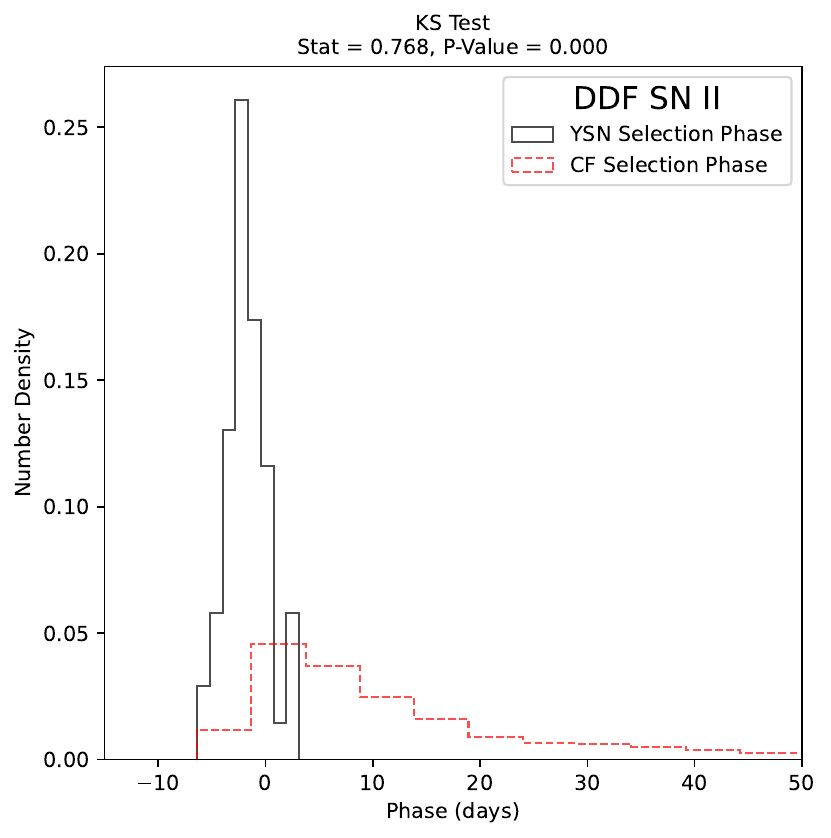}
\includegraphics[width=0.8\linewidth]{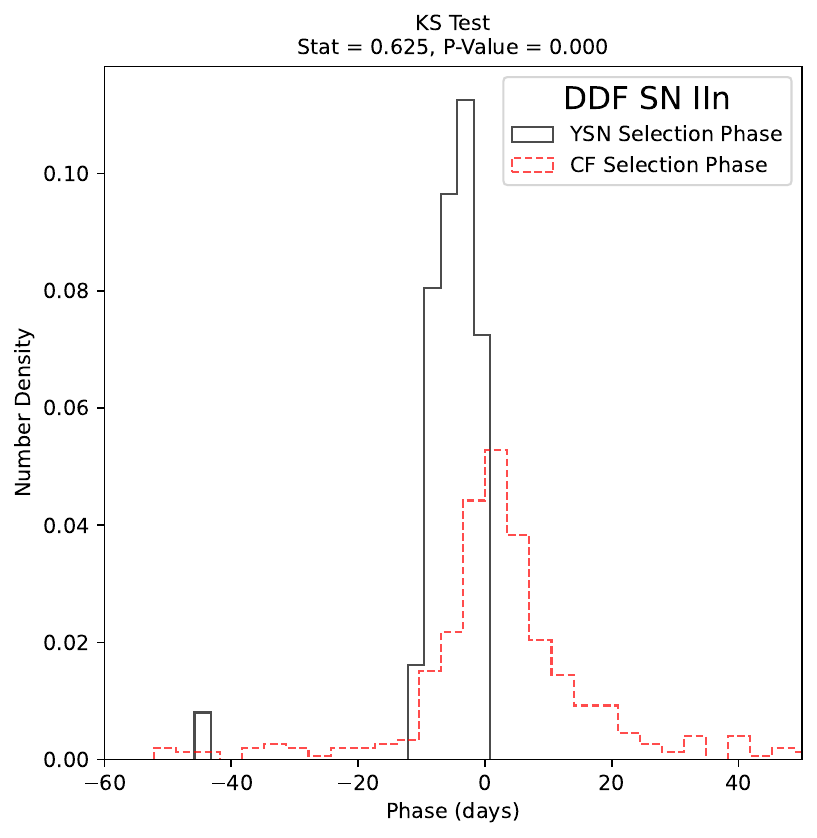}
\includegraphics[width=0.8\linewidth]{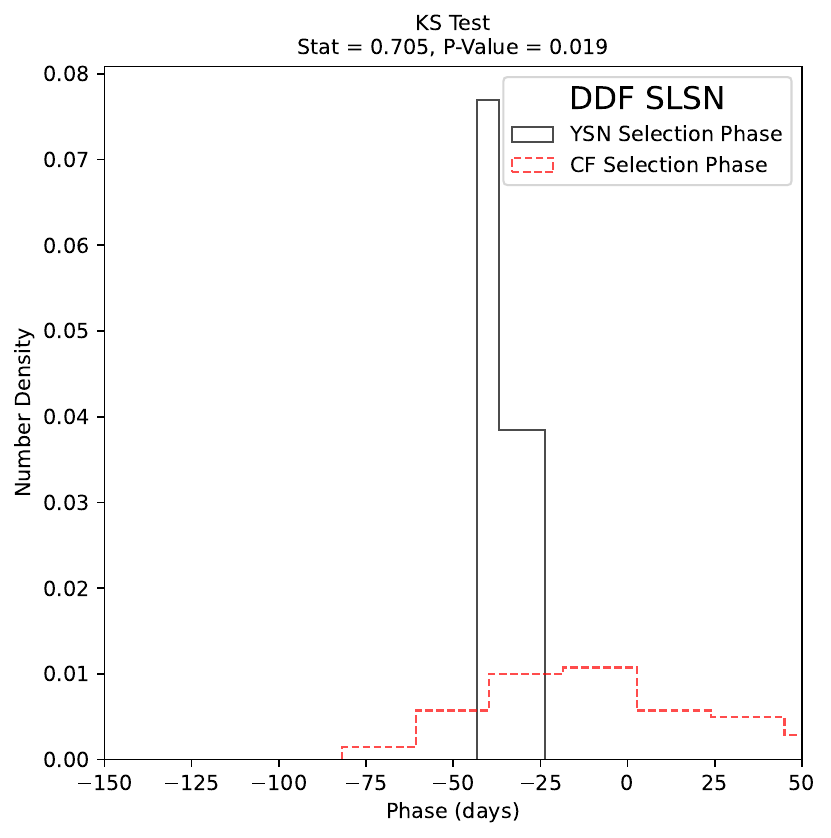}
\includegraphics[width=0.8\linewidth]{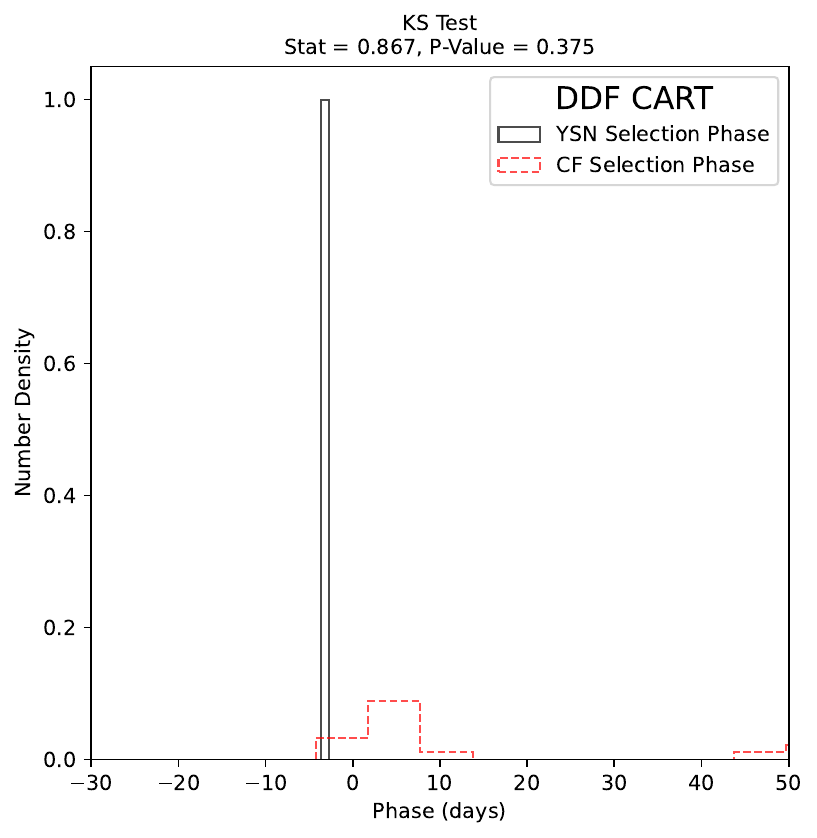}
\includegraphics[width=0.8\linewidth]{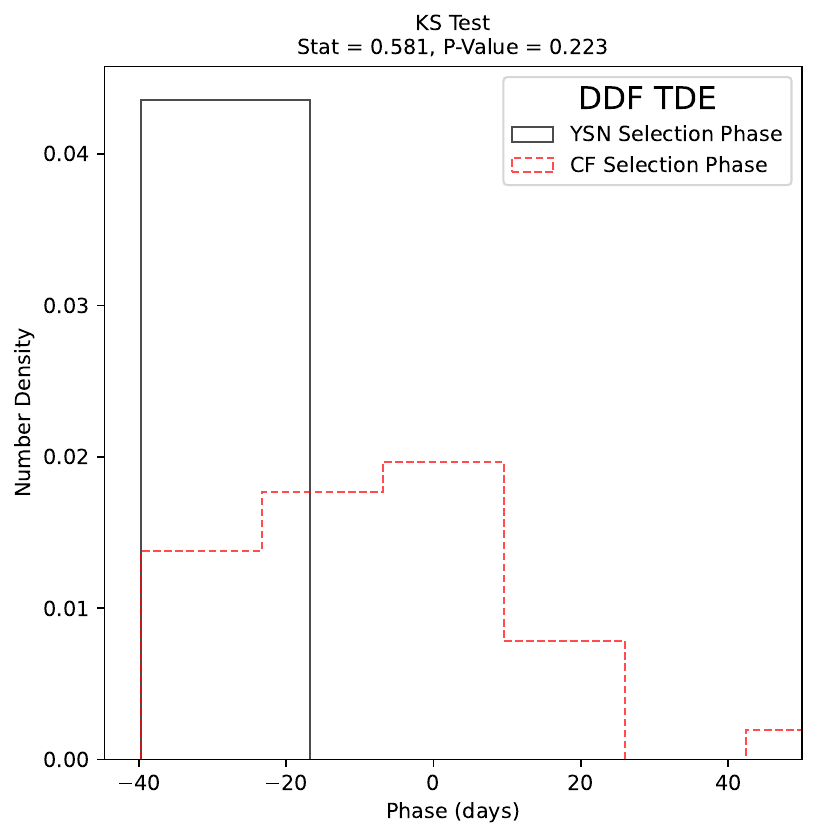}


\bsp	
\label{lastpage}
\end{document}